\documentclass{aa}
\usepackage{comment}
\usepackage{graphicx}
\usepackage{txfonts}
\usepackage{multirow}
\usepackage{subcaption}
\usepackage{adjustbox}
\usepackage{amssymb}
\usepackage{xcolor}
\usepackage{colortbl}
\usepackage[colorlinks=true,allcolors=blue]{hyperref}

\makeatletter
\renewcommand*\aa@pageof{, page \thepage{} of \pageref*{LastPage}}
\makeatother

\usepackage{textgreek}
\usepackage{xargs}
\usepackage{xcolor}
\usepackage{xspace}

\let\oldtextsigma\textsigma
\renewcommand{\textsigma}{\oldtextsigma\xspace}
\let\oldAA\AA
\renewcommand{\AA}{\text{\oldAA}\xspace}
\def\w80{\ensuremath{w_{80}}\xspace}

\newcommand{\mum}{\text{\textmu m}\xspace}

\newcommandx{\fluxdcgs}[1][1=-20]{$\times 10^{[#1]}$~erg~s$^{-1}$~cm$^{-2}$~\AA$^{-1}$\xspace}

\newcommand{\Halpha}{\text{H\textalpha}\xspace}
\newcommand{\Hbeta}{\text{H\textbeta}\xspace}

\newcommand{\Hdelta}{\text{H\textdelta}\xspace}
\newcommand{\Hzeta}{\text{H\textzeta}\xspace}
\newcommandx{\permittedEL}[6][1=O,2=III,3=,4=,5=,6=]{\text{{#1}\,{\sc{#2}}{#3}{#4}{#5}{#6}}\xspace}
\newcommandx{\semiforbiddenEL}[6][1=O,2=III,3=,4=,5=,6=]{\text{{#1}\,{\sc{#2}}]{#3}{#4}{#5}{#6}}\xspace}
\newcommandx{\forbiddenEL}[6][1=O,2=III,3=,4=,5=,6=]{\text{[{#1}\,{\sc{#2}}]{#3}{#4}{#5}{#6}}\xspace}

\newcommandx{\HII}{\permittedEL[H][ii]}
\newcommandx{\HeI}{\permittedEL[He][i]}
\newcommandx{\HeIL}[1][1=3889]{\permittedEL[He][i][\,\textlambda][#1]}
\newcommandx{\HeIIL}[1][1=4686]{\permittedEL[He][ii][\,\textlambda][#1]}
\newcommand{\OIII}{\forbiddenEL[O][iii]}
\newcommandx{\OIIIL}[1][1=5007]{\forbiddenEL[O][iii][\textlambda][#1]}
\newcommand{\OIIIall}{\forbiddenEL[O][iii][\textlambda][\textlambda][4959,][5007]}
\newcommandx{\NIL}[1]{\forbiddenEL[N][i][\textlambda][5200]}

\newcommand{\OII}{\forbiddenEL[O][ii]}
\newcommandx{\OIIL}[1][1=3727]{\forbiddenEL[O][ii][\textlambda][#1]}
\newcommand{\OIIall}{\forbiddenEL[O][ii][\textlambda][\textlambda][][3726,3729]}
\newcommand{\NeIIIall}{\forbiddenEL[Ne][iii][\textlambda][\textlambda][][3869,3967]}

\newcommand{\NeIII}{\forbiddenEL[Ne][iii][\textlambda][3869]}
\newcommand{\NII}{\forbiddenEL[N][ii][\textlambda][6584]}
\newcommandx{\NIIL}[1][1=6583]{\forbiddenEL[N][ii][\textlambda][#1]}
\newcommand{\NeIIIb}{\forbiddenEL[Ne][iii][\textlambda][3967]}
\newcommand{\OIIAuall}{\forbiddenEL[O][ii][\textlambda][\textlambda][7319--][7332]}
\newcommand{\OIall}{\forbiddenEL[O][i][\textlambda][\textlambda][6300,][6363]}
\newcommand{\NIall}{\forbiddenEL[N][i][\textlambda][\textlambda][5198,][5200]}

\newcommand{\jwst}{\textit{JWST}\xspace}

\graphicspath{{./}{figures/}}

\begin{document}
	
	\title{Spatially resolved emission lines in galaxies at $4\leq z < 10$ from the JADES survey: evidence for enhanced central star formation}
	
	\titlerunning{Spatially resolved emission line properties of galaxies at $4\leq z \leq 10$}
	\authorrunning{R. Tripodi et al.}
	
\author{Roberta~Tripodi
    \inst{1\!-\!6}\thanks{\email{roberta.tripodi@inaf.it}}
    \and Francesco~D'Eugenio
    \inst{5,6}
    \and Roberto~Maiolino
    \inst{5,6,7}
    \and Mirko Curti
    \inst{8}
    \and Jan~Scholtz
    \inst{5,6}
    \and Sandro~Tacchella
    \inst{5,6}
    \and Andrew~J.~Bunker
    \inst{9}
    \and James~A.~A.~Trussler
    \inst{10}
    \and Alex~J.~Cameron
    \inst{9}
    \and Santiago~Arribas
    \inst{11}
    \and William~M.~Baker
    \inst{5,6}
    \and Maruša Bradač
    \inst{1}
    \and Stefano~Carniani
    \inst{12}
    \and St{\'e}phane~Charlot
    \inst{13}
    \and Xihan~Ji
    \inst{5,6}
    \and Zhiyuan~Ji
    \inst{14}
    \and Brant~Robertson
    \inst{15}
    \and Hannah~\"Ubler
    \inst{5,6}
    \and Giacomo~Venturi
    \inst{12}
    \and Christopher~N.~A.~Willmer
    \inst{14}
    \and Joris~Witstok
    \inst{5,6}    
    }

 \institute{University of Ljubljana FMF, Jadranska 19, 1000 Ljubljana, Slovenia
        \and
        Dipartimento di Fisica, Università di Trieste, Sezione di Astronomia, Via G.B. Tiepolo 11, I-34131 Trieste, Italy
        \and
        INAF - Osservatorio Astronomico di Trieste, Via G. Tiepolo 11, I-34143 Trieste, Italy
         \and
         IFPU - Institute for Fundamental Physics of the Universe, via Beirut 2, I-34151 Trieste, Italy
         \and
        Cavendish Laboratory - Astrophysics Group, University of Cambridge, 19 JJ Thomson Avenue, Cambridge, CB3 0HE, UK
         \and
         Kavli Institute for Cosmology, University of Cambridge, Madingley Road, Cambridge CB3 0HA, UK
         \and 
         Department of Physics and Astronomy, University College London, Gower Street, London WC1E 6BT, UK
         \and 
         European Southern Observatory, Karl-Schwarzschild-Strasse 2, 85748 Garching, Germany
         \and
         Department of Physics, University of Oxford, Denys Wilkinson Building, Keble Road, Oxford OX1 3RH, UK
         \and
         Jodrell Bank Centre for Astrophysics, University of Manchester, Oxford Road, Manchester M13 9PL, UK
         \and 
         Centro de Astrobiolog\'ia (CAB), CSIC–INTA, Cra. de Ajalvir Km.~4, 28850- Torrej\'on de Ardoz, Madrid, Spain
         \and
         Scuola Normale Superiore, Piazza dei Cavalieri 7, I-56126 Pisa, Italy
         \and
         Sorbonne Universit\'e, CNRS, UMR 7095, Institut d'Astrophysique de Paris, 98 bis bd Arago, 75014 Paris, France
         \and
        Steward Observatory, University of Arizona, 933 North Cherry Avenue, Tucson, AZ 85721, USA
        \and
        Department of Astronomy and Astrophysics, University of California, Santa Cruz, 1156 High Street, Santa Cruz, CA 95064, USA
         \\
             }

	\abstract{
		We present the first statistical investigation of spatially resolved emission-line properties in a sample of 63 low-mass galaxies at $4\leq z<10$, using \jwst/NIRSpec MSA data from the \jwst Advanced Deep Extragalactic (JADES) survey focusing on deep, spatially resolved spectroscopy in the GOODS-S extragalactic field. 
		By performing a stacking of the 2D spectra of the galaxies in our sample, we find an increasing or flat radial trend with increasing radius for \OIIIL/\Hbeta and a decreasing one for \NeIII/\OIIL (3--4 $\sigma$ significance). These results are still valid when stacking the sample in two redshift bins (i.e., $4\leq z<5.5$ and $5.5\leq z<10$). The comparison with star-formation photoionization models suggests that the ionization parameter increases by $\sim 0.5$ dex with redshift. We find a tentative metallicity gradient that increases with radius (i.e., `inverted') in both redshift bins.
		Moreover, our analysis reveals strong negative gradients for the equivalent width of \Hbeta (7$\sigma$ significance).
		This trend persists even after removing known AGN candidates, therefore, it is consistent with a radial gradient primarily in stellar age and secondarily in metallicity.
		Taken all together, our results suggest that the sample is dominated by active central star formation, with possibly inverted metallicity gradients sustained by recent episodes of accretion of pristine gas or strong radial flows.
		Deeper observations and larger samples are needed to confirm these preliminary results and to validate our interpretation.
	}
	
	
	\maketitle
	
	\nolinenumbers

	\section{Introduction}
	
	The distribution of metals within galaxies is a consequence of the processes of metal production, circulation, and dilution, the latter primarily attributed to the phenomenon of accretion. Therefore, studying metallicity gradients is crucial for understanding how these physical processes work.
	Mostly due to selection effects, the majority of high-redshift galaxies ($z>4$) are likely caught in a complex and turbulent phase of their evolution, characterized by gas accretion, frequent merging events, and by gas outflows due to stars and/or active galactic nuclei (AGNs) feedback \citep[see e.g.][]{bischetti2019,bischetti2021,tripodi2024b,neeleman2021,shao2019,dayal2018}. These mechanisms clearly affect the content and distribution of heavy elements in the galaxy's interstellar medium \citep[ISM, ][]{dave2011}. In the local Universe, the spatial distribution of metals in galaxies has been studied precisely, assessing the presence of radial variations and chemical enrichment levels across the galaxy. A large number of local star-forming (SF) galaxies exhibit negative  radial gradients in their gas-phase metallicity (or simply metallicity), with the inner regions more chemically enriched with respect to the outskirts \citep[e.g., ][]{magrini2010, berg2015, belfiore2017,bresolin2016,li2018}. These gradients are generally interpreted as being indicative of the so-called inside-out growth scenario of galaxy formation \citep{samland1997,portinari1999,gibson2013,pezzulli2017}. Deviations from negative gradients provide specific information on the evolutionary phase of galaxies. The observation of flattening gradients beyond a certain radius may result from radial mixing processes \citep[e.g. outflow; see e.g. ][]{choi2020}, or (re)accretion of metal-enriched gas in the outer regions \citep{bresolin2012}, or an ongoing merger \citep{kewley2010,rupke2010a,rupke2010b}. Positive (or inverted, i.e. increasing with radius) gradients may indicate a high rate of accretion of pristine gas \citep{sanchez2018}. Moving from the local Universe to cosmic noon, it has been shown that galaxies at $1<z<2$ also grew inside-out, similar to their local counterparts \citep{nelson2016,suzuki2019,tacchella2015,tacchella2018,wang2019,wang2020,wang2022} . However, metallicity gradients seem to evolve \citep[see e.g.,][]{li2022}. \citet{curti2020}  investigated the metallicity gradients of a sample of 42 galaxies between $1.2<z<2.5$,
	finding that $\sim 85\%$ of their galaxies are characterized by metallicity gradients shallower than 0.05 dex kpc$^{-1}$, and $\sim 89\%$ are consistent with a flat slope within $3\sigma$, suggesting a mild evolution with cosmic time. They also discovered three galaxies with inverted gradients, suggesting recent episodes of pristine gas accretion or strong radial flows.
	
	The advent of \jwst enabled for the first time the study of metallicity gradients at $z>3\text{--}5$, with the same rest-frame optical diagnostics used for galaxies at $0<z<2$ \citep[for studies of galaxies at $z<5$ see e.g., ][]{troncoso2014}. However, current studies are limited to a couple of bright, extended systems that are not representative of the galaxy population at $z>3$ \citep[see studies performed using integral field unit (IFU) or NIRISS observations or modelling FIR emission lines from ALMA, e.g.,][]{rodriguez-delpino+2023,arribas+2023,venturi2024,vallini2024,wang2022}. The metallicity gradients of the bulk of galaxies at redshifts higher than $z>4$ are still poorly understood. Being able to measure the metallicity at early epochs gives us the invaluable opportunity to study the history of the baryonic cycle and its influence on the evolution of galaxies. Moreover, at very high-z the interpretation of radial measurements of emission line ratios is more challenging than at low-z, since high-z galaxies show irregular shapes with several (merging) components and/or clumps. 
	
	The collisionally excited \OIIIL[5007] and the recombination \Hbeta optical emission lines are commonly used as diagnostics for the properties of gas in star-forming regions. In particular, \Hbeta is driven primarily by ionizing radiation, while \OIIIL[5007] is sensitive to the gas-phase metallicity and the ionization parameter. In combination with other lines (e.g., \NeIII and \OIIall), these two diagnostics allow us to measure the metallicity, electron temperature, ionization parameter, and hardness of the radiation field in the ISM of a star-forming galaxy \citep{osterbrock1989}. Indeed, a powerful way to trace metallicity gradients at high-z is by comparing the \OIIIL[5007]/\Hbeta emission line ratio with the ratio between $\NeIII$ and $\OIIall$ (hereafter $\OIIL[3727]$) emission lines\footnote{Ionization potentials for \permittedEL[O][i], \permittedEL[O][ii] and \permittedEL[Ne][ii] are 13.61 eV, 35.12 eV and 40.96 eV, respectively.} \citep[e.g.,][]{nagao2006, levesque2014}. Neon is produced during the late evolutionary stages of massive stars and it is expected to closely track oxygen abundance \citep{thielemann1994,henry1999}. Along with oxygen, neon is one of the principal coolants in $\HII$ regions. Moreover, $\NeIII$ arises from a broader range of regions in the ionized nebula when compared to $\OII$, given that $\NeIII$'s high critical density ($\log n[{\rm cm^{-3}}]=7$, see \citealt{appenzeller1988}) makes its flux insensitive to the electron density even in high-density regions. The $\NeIII/\OIIL[3727]$ ratio has been proven a better diagnostic of ionization parameter than $\OIIIL[5007]/\OIIL[3727]$, 
	with a greater sensitivity at shorter wavelengths that accommodates more of the ionizing photons produced by young massive stars. It is also insensitive to reddening effects and usable as an empirical diagnostic of ionization parameter out to higher redshift than $\OIIIL[5007]/\OIIL[3727]$, since $\OIIIL[5007]$ is emitted at longer wavelength than $\NeIII$. The degeneracy between metallicity and ionization parameter has indeed proven challenging to disentangle when trying to calibrate abundance diagnostics for SF galaxies and $\HII$ regions. Therefore, when investigating metallicity, it is indispensable to have a tracer that is also able to constrain the ionization parameter \citep[e.g., ][]{mcgaugh1991, kewley2002}.
	
	Additionally, \Hbeta is particularly powerful in probing ionized regions around young, massive stars. Leaving dust extinction aside, the intensity of the \Hbeta line is directly linked to the ionizing photons emitted by these stars, providing a valuable metric for quantifying the ongoing star-forming activity. As a SFR diagnostic, \Hbeta, like all the Balmer lines, inherits the same strengths and weaknesses of \Halpha: it is equally sensitive to variations in the IMF and to absorption of Lyman-continuum photons by dust within star-forming regions \citep{moustakas2006}. More precisely, assuming a star-forming origin, the equivalent width (EW) of \Hbeta is a tracer of the specific SFR (sSFR) in star-forming galaxies, modulo the absorption of ionizing photons by dust in HII regions. 
	
	In this work, we aim to investigate the emission-line gradients in a sample of 63 galaxies at $4\leq z<10$ from the JADES survey. Specifically, we focus on spatially resolved analyses of the \OIIIL[5007]/\Hbeta and $\NeIII/\OIIL[3727]$ ratios to investigate the distribution of metallicity in high-z galaxies, and of the EW$_{\rm \Hbeta}$ as a tracer of the sSFR. We focus on the EW and these specific line ratios as, given that they are very close in wavelength, they do not suffer severely of the wavelength dependent PSF. Although it would be tempting to consider line ratios involving other lines more widely separated in wavelength, the wavelength dependent PSF results in blending different regions at different wavelength and different slit losses. Selecting diagnostics consisting in lines that are close in wavelength also mitigate dust reddening effects.
	
	This is the first attempt to perform spatially resolved metallicity studies at $z>4$ in a large sample of galaxies. The paper is structured as follows. In Sect. \ref{sec:obs}, we present the observations and the data reduction; in Sect. \ref{sec:analysis}, we report the analysis and results on stacked samples; in Sect. \ref{sec:disc}, we discuss the implications of our findings, comparing our results with observations at different redshifts and with photoionization models. Throughout the paper, we adopt a $\Lambda$CDM cosmology from \citet{planck2018}: $H_0=67.4\ \rm km\ s^{-1}\ Mpc^{-1}$, $\Omega_m = 0.315$ and $\Omega_{\Lambda} = 0.685$. We assume everywhere a Chabrier initial mass function \citep{chabrier2003}. All physical distances are proper distances.

	\section{Observations and data reduction}
	\label{sec:obs}
	
	We use publicly available \jwst/NIRSpec data from the \jwst Advanced Deep Extragalactic Survey \citep[\href{https://jades-survey.github.io/}{JADES}, ][]{eisenstein+2023, bunker+2023b, rieke2023}, a collaboration between the \jwst/NIRCam and NIRSpec GTO teams. In this work, we focus on deep, spatially resolved spectroscopy in the GOODS-S extragalactic field \citep{giavalisco+2004}. Specifically, we analyse 63 galaxies at $4\leq z <10$. These data were obtained from 
	Programme ID 1210 (PI: N.~L\"utzgendorf; henceforth: PID), and used the NIRSpec micro-shutter assembly \citep[MSA;][]{jakobsen+2022,ferruit2022}. For the galaxy images, we used publicly available NIRCam data from JADES itself (PID~1180; PI: D.~Eisenstein), from the \jwst Extragalactic Medium-band Survey (JEMS, PID~1963, PIs: C.\ C.\ Williams, S.\ Tacchella and M.\ Maseda; \citealp{williams2023}), and from the First Reionization Epoch Spectroscopic COmplete Survey (FRESCO, PID~1895; PI: P.~Oesch; \citealp{oesch2023}).
	
	Even though JADES includes observations with a range of dispersers 
	\citep{bunker+2023b, carniani+2023}, in this work we use only the PRISM/CLEAR observations, because they are the deepest available while still providing sufficient spectral resolution for our purpose of separating \OIIL from $\NeIII$, and \Hbeta from $\OIIIall$ at redshifts $z>4$. These data were observed using
	a 3-shutter slit, with 1-shutter nodding to provide accurate background subtraction.
	The observations also included dithering to explore different regions of the detector and to safeguard against `disobedient' shutters, i.e., shutters that are in a different open or closed state than the requested configuration. The observations consist of three pointings, where each target is allocated to up to three pointings according to its priority \citep[as described in ][]{bunker+2023b}. The optimisation of the MSA allocation was performed using the \textsc{EMPT} software \citep{bonaventura+2023}.
	For our targets, the exposure times range from 9.3 to 28~hours.
	
	The data reduction is described in \citet{bunker+2023b}; we report here only the most relevant steps. The data reduction pipeline is based on the ESA NIRSpec Science Operations Team pipeline \citep{alvesdeoliveira2018, ferruit2022}, and will be described in Carniani et al. (in~prep.). Background subtraction was performed by using the local background from adjacent shutters, but extended sources and sources with contaminants were pre-identified by visual inspection and were not self subtracted, by using only empty shutters for the background. Some degree of self subtraction may still be present for the most extended sources (i.e., those extending more than two shutters), 
	but these are only relevant at the lowest redshifts and do not affect the sample we use in this paper.
	The wavelength calibration includes a correction for intra-shutter
	target position. The pipeline also applies a wavelength-dependent path-loss correction, to  account for flux falling outside the micro-shutters. This takes into account
	the wavelength-dependent size of the NIRSpec PSF as well as the intra-shutter target position, assuming a point-source light distribution. While this assumption is incorrect at lower redshifts, it is an excellent approximation for the redshift range used in this paper ($z>4$). Because the nominal spectral resolution of the prism varies by an order of magnitude \citep[$R=30\text{--}330$ over the wavelength range,][]{jakobsen+2022}, the spectra are binned on an irregular grid, with varying pixel size matching the spectral resolution and ensuring Nyquist sampling of the resolution element.
	

	\section{Analysis and Results}
	\label{sec:analysis}
	
	\subsection{Line fitting procedure}
	\label{sec:fitting}
	
	The main goal of this work is to study the line properties of $\OIIIL[5007],\Hbeta,\NeIII$ and $\OIIL[3727]$. Therefore, in this section, we briefly explain the procedure adopted for fitting the 2D spectra of our targets. 
	
	We fit separately in pairs \OIIIL[5007] and \Hbeta, and \OIIL[3727] and \NeIII, with their corresponding underlying continua. We model the emission lines with Gaussian functions and the two underlying continua with 1\textsuperscript{st}-order polynomials\footnote{Since the continuum is fitted just in proximity of each pair of lines, a 1\textsuperscript{st}-order polynomial is enough to capture the shape of the continuum, and it leads to the same results as if considering a power-law functional form.}. The doublet $\OIIIall$ has been fitted fixing the ratio between the peak fluxes (peak$_{\OIIIL[4959]}/{\rm peak}_{\OIIIL[5007]}=0.335$) and the wavelength separation ($\Delta \lambda = 47.94$ \AA) of the two emission lines, and using the same FWHM for both emission lines. Similarly, we fit simultaneously the $\NeIII$ and $\NeIIIb$ doublet, adopting a ratio of 0.301 between the latter and the former, and a rest-frame wavelength separation of $98.73$ \AA, and considering the same FWHM for both emission lines. The doublet \OIIL is always blended given the resolution of our spectra, so we fitted it with a single Gaussian. This reduces to 3 the number of free parameters for both the $\OIII$ and the $\NeIII$ doublets. Therefore, we have 8 free parameters for each pair of lines (i.e., peak flux, peak wavelength, FWHM for \OIIIL[5007] and same for \Hbeta, slope and intercept for their underlying continuum; same for \OIIL[3727] and \NeIII). We explore the 8-dimensional parameter space for each pair of lines using a Markov chain Monte Carlo (MCMC) algorithm implemented in the \texttt{EMCEE} package \citep{foreman2013}, assuming uniform priors for the fitting parameters, considering 10 walkers per parameter and 1500 trials (the typical burn-in phase is $\sim$ 500 trials). Analogously, we fit the \OIIIL[5007], \Hbeta, \OIIL[3727] and \NeIII emission lines and corresponding continua in the 5-pixel boxcar extraction of the 1D spectrum\footnote{As described in \citet{bunker+2023b}, the 5-pixel boxcar extraction is not performed on the combined 2D spectra, but on each individual exposure; the resulting 1D spectra are then combined in the final 1D spectrum.}. Finally, we compute the integrated fluxes by integrating the best-fitting functions for each emission line.
	
	Moreover, regarding the 2D spectra, we measure the EW of \Hbeta for each trial in the chain. Then, we derive the best-fitting value from the 50\textsuperscript{th} percentile of the EW$_{\rm \Hbeta}$'s chain, and the error on EW$_{\rm \Hbeta}$ from the 16\textsuperscript{th} and 84\textsuperscript{th} percentiles.  
	
	
	\subsection{Line blending and contamination correction}
	\label{sec:contam}
	
	When measuring \NeIII, we also include flux from \Hzeta ($\lambda=3890.17$~\AA) and \HeIL, both of  which are  blended with \NeIII at the resolution of the prism. To estimate the contamination, we proceed as follows. 
	For \Hzeta, we calculate the ratio \Hdelta/(\NeIII + \Hzeta + \HeIL) from the stacked 1D spectrum (see Sect.~\ref{sec:stacking}), finding a value of $0.35\pm0.01$. This means that the ratio \Hzeta/(\NeIII + \Hzeta + \HeIL) will be less than 0.14, assuming the Balmer ratios from Case~B recombination,
	$T_\mathrm{e}=10,000$~K and $n_\mathrm{e}=100~\mathrm{cm}^{-3}$ \citep{storey+hummer1995}. We apply no reddening correction to
	\Hdelta/\NeIII, therefore this contamination fraction is strictly an upper limit. For \HeIL, the estimate is much more uncertain, because this line can be optically thick \citep[e.g.,][]{robbins1968}.
	As a tentative estimate, we use the prism spectra to measure the \HeIL[5877] line and then use models to infer the \HeIL contamination to \NeIII. For $n_\mathrm{e} = 100~\mathrm{cm}^{-3}$ and temperatures
	$T_\mathrm{e}=5,000$, 10,000 and 20,000~K, \citet{benjamin+1999} gives \HeIL/\HeIL[5877] of 0.6, 0.8 and 1.1, respectively (these are intrinsic values, without reddening correction). From the stacked 1D spectrum, we also measure a ratio of $\HeIL[5877]/\NeIII=0.23\pm0.01$, thus even assuming the highest \HeIL/\HeIL[5877] value of 1.1, the overall contamination should be of order 0.25, similar to what we estimated for \Hzeta. Thus we have
	both $\Hzeta / (\NeIII + \Hzeta + \HeIL) < 0.14$ and $\HeIL / (\NeIII + \Hzeta + \HeIL) < 0.25$. Adding these two constraints, we obtain $(\Hzeta+\HeIL)/(\NeIII+\Hzeta+\HeIL)<0.39$, implying $\NeIII/(\NeIII + \Hzeta + \HeIL)>0.61$. This is an upper limit on the
	contamination, because, as we noted, we did not apply a reddening correction for \Hzeta/\Hdelta or \HeIL/\HeIL[5877]. In addition, it has to be noted that \HeI models are still quite uncertain, even for optically thin lines \citep{benjamin+1999} -- let alone the optically-thick \HeIL. In light of these difficulties, we do not apply the estimated correction. However, we show its magnitude in the relevant figures.
	
	\begin{figure*}
		\centering
		\includegraphics[width=\linewidth]{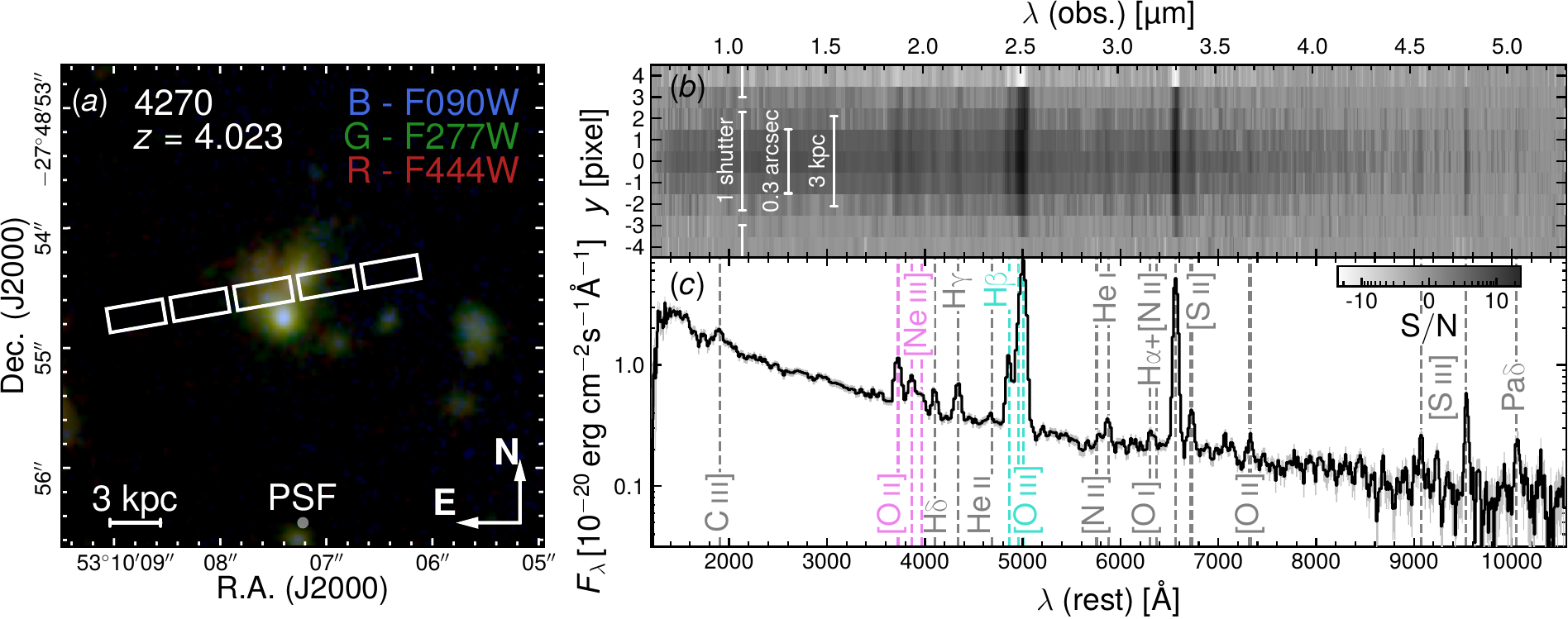}
		{\phantomsubcaption\label{fig:2d-spec-alt.a}
			\phantomsubcaption\label{fig:2d-spec-alt.b}
			\phantomsubcaption\label{fig:2d-spec-alt.c}}
		\caption{(a): 3-arcsec false-colour cutout. The nominal location of the shutters is overlaid; the MSA acquisition accuracy is better than one NIRCam pixel (0.03 arcsec); the gray circle represents the FWHM NIRSpec PSF from \texttt{webbpsf}, at the observed wavelength of \OIIIL. (b): 2D spectrum of 4270 at $z=4.023$. The y axis has been shifted so that the centre of the galaxy is at pixel 0. The physical scale of one pixel is reported on the plot (0.1 arcsec/pixel). (c): 1D spectrum of 4270. Note the complex morphology of this galaxy: the synergy between spectroscopy and imaging is vital for interpreting correctly spatially resolved slit spectroscopy. The simultaneous presence of high-ionisation and low-ionisation species, including auroral lines (cf.~\HeIIL vs. \NIIL[5755], \OIall, \OIIAuall, possibly \NIall) may indicate a complex interplay between AGN photoionization and shocks.} 
	\label{fig:2d-spec-alt}
\end{figure*}
\begin{figure*}
	\centering
	\includegraphics[width=\linewidth]{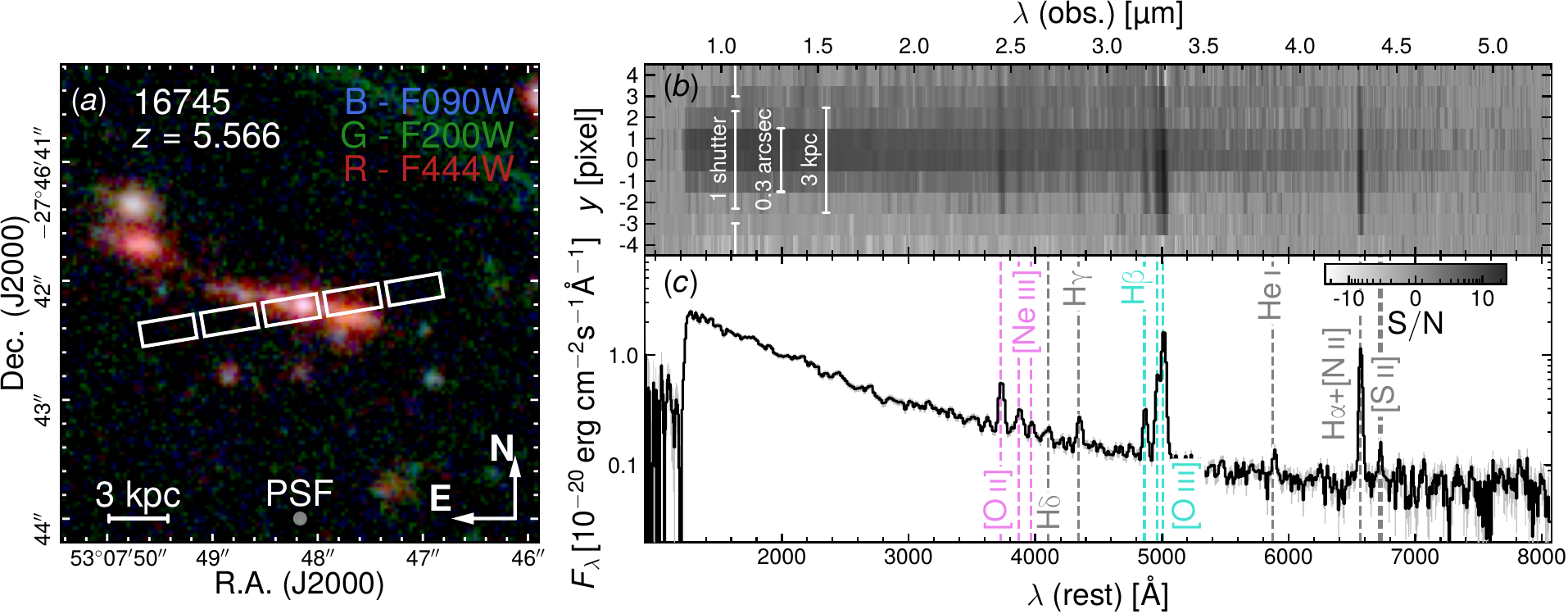}
	{\phantomsubcaption\label{fig:2d-spec.a}
		\phantomsubcaption\label{fig:2d-spec.b}
		\phantomsubcaption\label{fig:2d-spec.c}}
	\caption{Same as Fig. \ref{fig:2d-spec-alt} for 16745 at $z=5.566$.}
	\label{fig:2d-spec}
\end{figure*}

\subsection{Line ratios: stacking analysis}
\label{sec:stacking}

We aim to perform a spatially resolved study of the \OIIIL[5007]/\Hbeta and \NeIII/\OIIL[3727] line ratios in a sample of galaxies at $4\leq z<10$. Therefore, we initially select all the publicly available galaxies belonging to the JADES survey with S/N$\geqslant 5$ on the flux of each of the four spectral lines\footnote{These fluxes are also publicly available, and are computed from the line fitting of the 1D spectrum of each source \citep{bunker+2023b}.} at $4\leq z<10$: we find that 19 objects satisfy these criteria. In Figs.~\ref{fig:2d-spec-alt} and~\ref{fig:2d-spec}, we show 3-arcsec false-colour cutouts and spectra of the two galaxies with the most extended and bright \OIIIL[5007], \Hbeta, \NeIII, and \OIIL[3727] emissions. We extract the spectra from 7 pixels (one spectrum for each pixel's row from -3 to +3, with pixel scale = 0.1 arcsec/pixel\footnote{At the median redshift of the three samples studied in this work (i.e., z=5.5, z=4.7, z=6), the pixel physical scale is 0.61 kpc, 0.66 kpc and 0.58 kpc, respectively.}; see e.g., 2D spectrum in Fig. \ref{fig:2d-spec-alt}) centred on the peak flux of the emission lines of each galaxy. We do not find any galaxy showing extended emission beyond pixel +3/-3. Considering that the spatial resolution of our data is about 2 pixels (see Appendix \ref{app:PSF} for estimate of the PSF size), we are not able to resolve emission within 1 pixel from the galaxy centre, that is within pixel -1 and 1.
We obtain detections or upper/lower limits on the \OIIIL[5007]/\Hbeta and/or \NeIII/\OIIL[3727] line ratios with a separation of more than 1 pixel from the galaxy centre only for 5 sources. They show flat radial trends of the line ratios. However, the complex morphology shown in the NIRCam images makes the interpretation of these tentative gradients arduous. Details on the analysis and results of individual sources can be found in Appendix \ref{app:indiv}.

\begin{figure}
	\centering
	\includegraphics[width=0.96\linewidth]{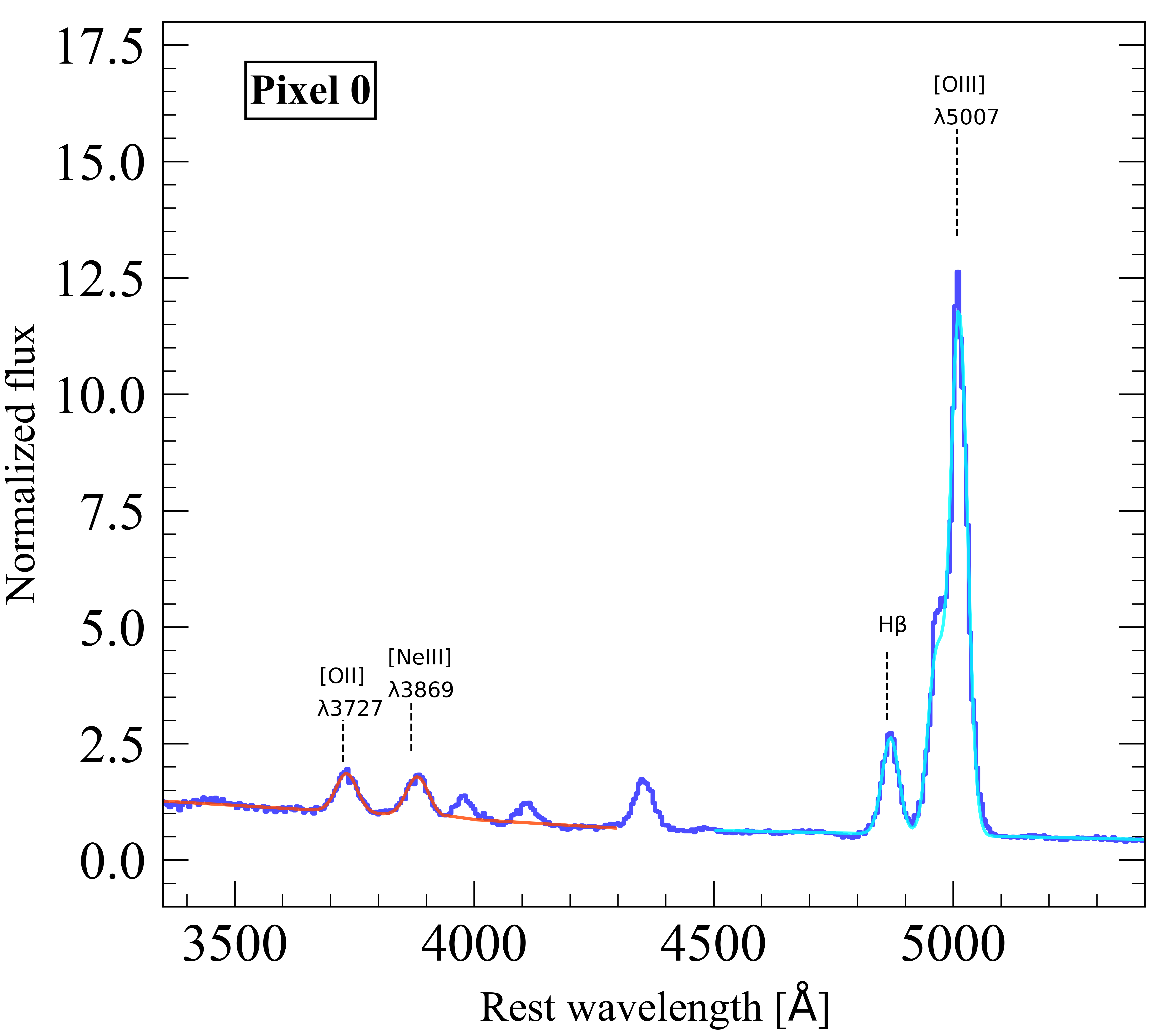}\\
	\includegraphics[width=0.92\linewidth]{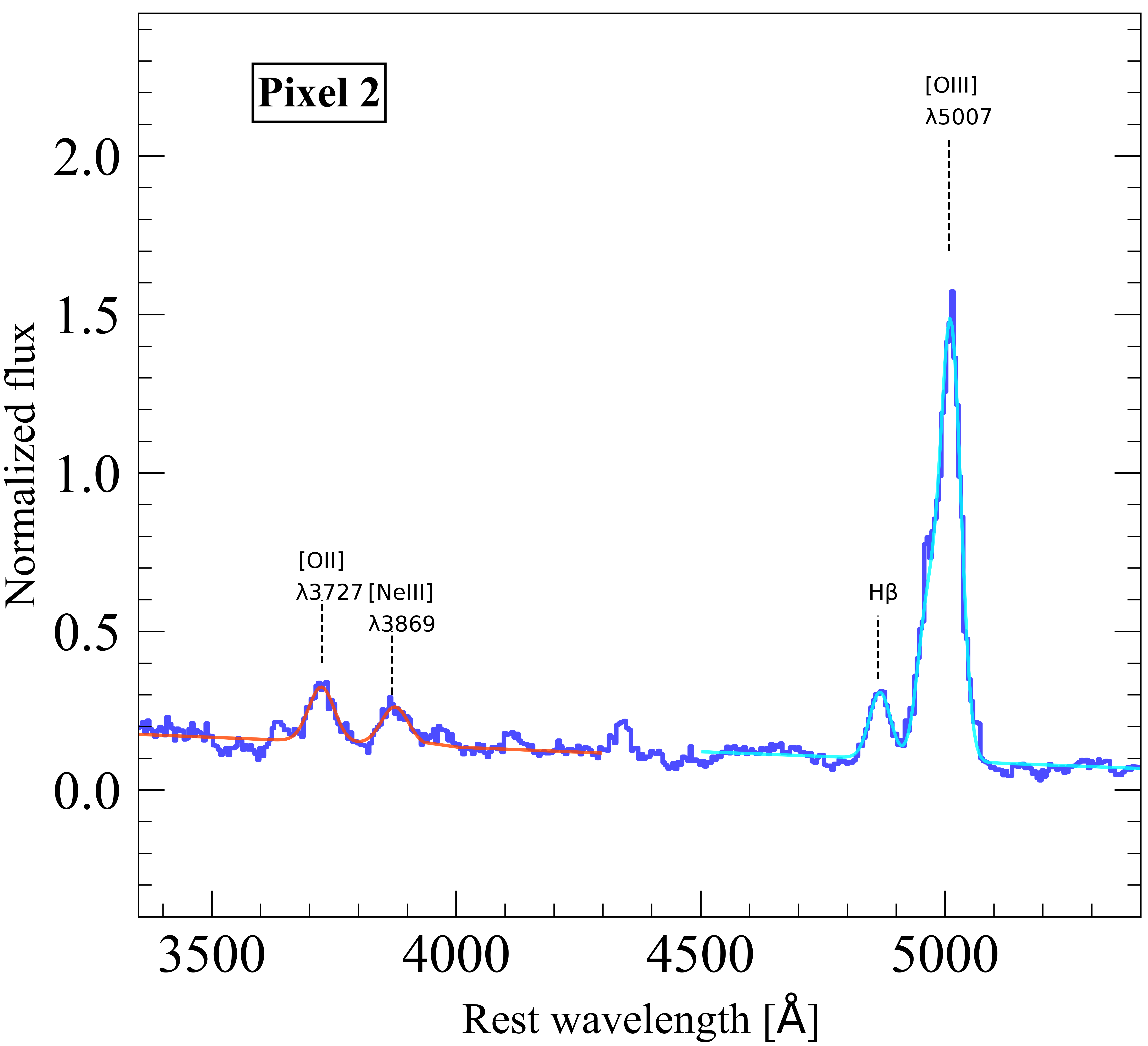}
	\caption{Stacked spectra using the NIRSpec/MSA prism spectra of galaxies at $4 \leq z< 10$ belonging to the JADES survey. Top panel: spectrum extracted from the central pixel of the 2D stacked spectrum. The cyan (violet) line is the best-fit function for the continuum and $\OIIIL[5007]$, \Hbeta ($\OIIL[3727], \NeIII$) spectral lines. Bottom panel: spectrum extracted from two pixel off central (i.e., $\sim 1.2$ kpc radius) of the 2D stacked spectrum. Colour coding and labels are the same as top panel. Even from a visual inspection of these two panels, one can notice the decrease of the \NeIII/\OIIL[3727] ratio towards larger radii, i.e., from pixel 0 (\NeIII/\OIIL[3727]$\sim 1$) to pixel 2 (\NeIII/\OIIL[3727]$< 1$).}
	\label{fig:spectra}
\end{figure}

Given the tentative results on the 19 individual sources, to increase the S/N of the emission lines, we stack the 2D and 1D spectra of all the sources at $4\leq z<10$ that have S/N$\geqslant 5$ on \OIIIL[5007]. This sample consists of 63 galaxies. To avoid being biased by the brightest objects in the sample, we start by dividing each spectrum by the integrated \OIIIL[5007] flux (for both the 1D and 2D analyses). We then weight the normalized spectra by the inverse of the variance\footnote{We checked that, for each galaxy spectrum, considering the variance either at a fixed wavelength or varying with wavelength did not alter our results for determining line ratios of \OIIIL[5007]/\Hbeta and \NeIII/\OIIL[3727], and EW$_{\Hbeta}$ on the stacked spectra. Not only the results obtained in these two ways are in agreement within the uncertainties, but also gradients and radial profiles are obtained at the same statistical significance.}. In addition to the full stack at $4<z<10$, we also consider two separate subsets, at $4\leq z < 5.5$ ($z_{\rm med}=4.7$) and at $5.5\leq z < 10$ ($z_{\rm med}=6$). These redshift bins are chosen to ensure a comparable number of objects in the two bins (30 and 33, respectively), and to cover time intervals of similar duration (0.50 and 0.57~Gyr, respectively). As an example, in Fig. \ref{fig:spectra}, we present the stacked spectra extracted from pixel 0 (top panel) and pixel 2 (bottom panel, corresponding to an average radial distance of $\sim 1.2$ kpc) of the 2D full stacked spectrum. The fit to the emission lines of interest 
has been performed as explained in Sect. \ref{sec:fitting}. 

\subsection{Radial gradients of emission-line ratios}

\begin{figure*}
	\centering
	\includegraphics[width=1\linewidth]{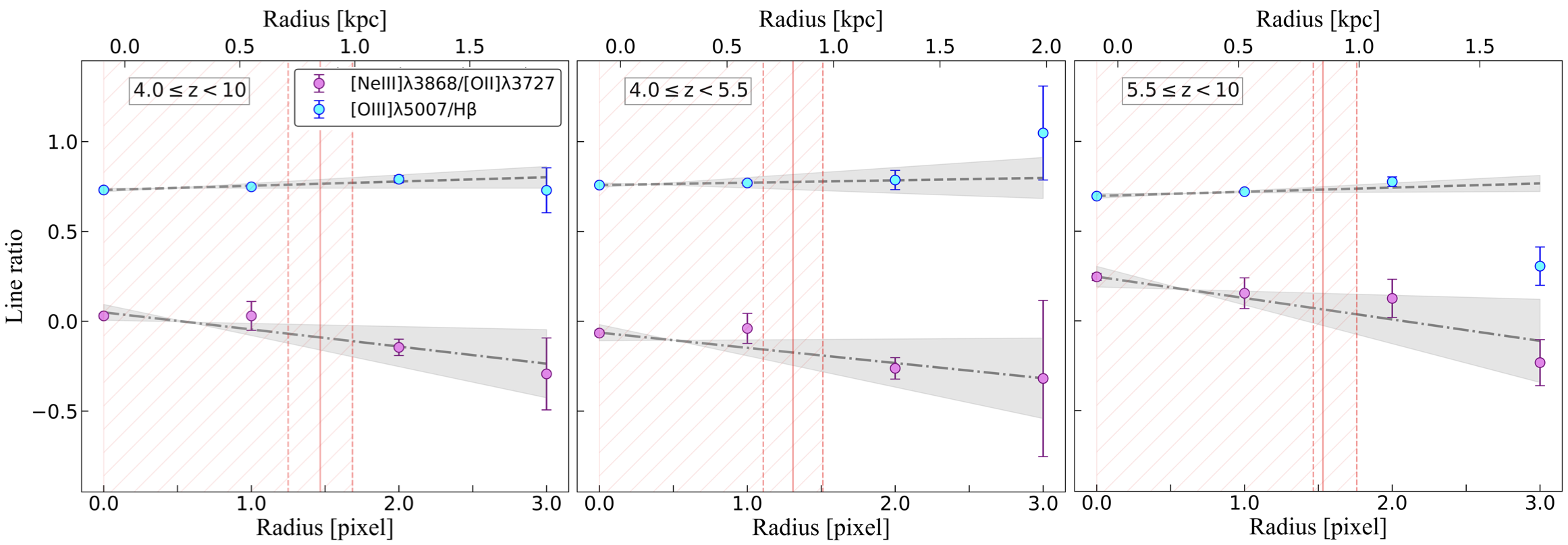}
	\caption{Averaged radial profiles of line ratios from the 2D stacked spectra at different redshift ranges. Cyan (violet) dots mark the logarithm of $\OIIIL[5007]$/\Hbeta ($\NeIII/\OIIL[3727]$) ratio as a function of radius expressed in pixels (pixel 0 is the centre of the galaxy). The left panel show the results stacking all galaxies at $4\leq z<10$; the central panel stacking all galaxies at $4\leq z<5.5$; the right panel stacking all galaxies at $5.5\leq z<10$. The fit for the $\OIIIL[5007]$/\Hbeta ($\NeIII/\OIIL[3727]$) profile is shown as a dashed (dash-dotted) line with shadowed region ($3\sigma$ confidence interval). The orange vertical lines mark the spatial resolution of the observations given by the 50th (solid), 16th (dashed left) and 84th (dashed right) percentiles of the distribution of the PSF size, computed from the redshift distribution of galaxies in each sample. For each panel, top x-axis is computed considering the median redshift of each sample that is: $z=5.5$ for the total stack, $z=4.7$ for the low-z stack, and $z=6$ for the high-z stack. These correspond to a scale of 6.113 kpc/", 6.614 kpc/" and 5.832 kpc/", respectively.}
	\label{fig:radial-prof}
\end{figure*}

The averaged radial profiles of line ratios in the three samples are shown in Fig. \ref{fig:radial-prof}: the errors on the averaged results account for both the statistical error of each measurement and the standard deviation between the measurements at each radius. We use a simple linear regression to quantify the gradient of these line ratios. The slopes for each sample and line ratio are reported in Tab. \ref{tab:grad}. While $\NeIII/\OIIL[3727]$ shows a decreasing trend towards the outskirts in all the samples, the $\OIIIL[5007]/\Hbeta$ ratio is almost constant with radius.
The measured gradients are only tentative, with a statistical significance between 3 and 4 $\sigma$; to confirm a real trend we would need deeper observations that allow to reduce the uncertainties and possibly to extend the detections to larger radii. We also note that the average value of both line ratios varies with redshift comparing the two redshift-bin sub-samples. 

\begin{figure*}
	\centering
	\includegraphics[width=1\linewidth]{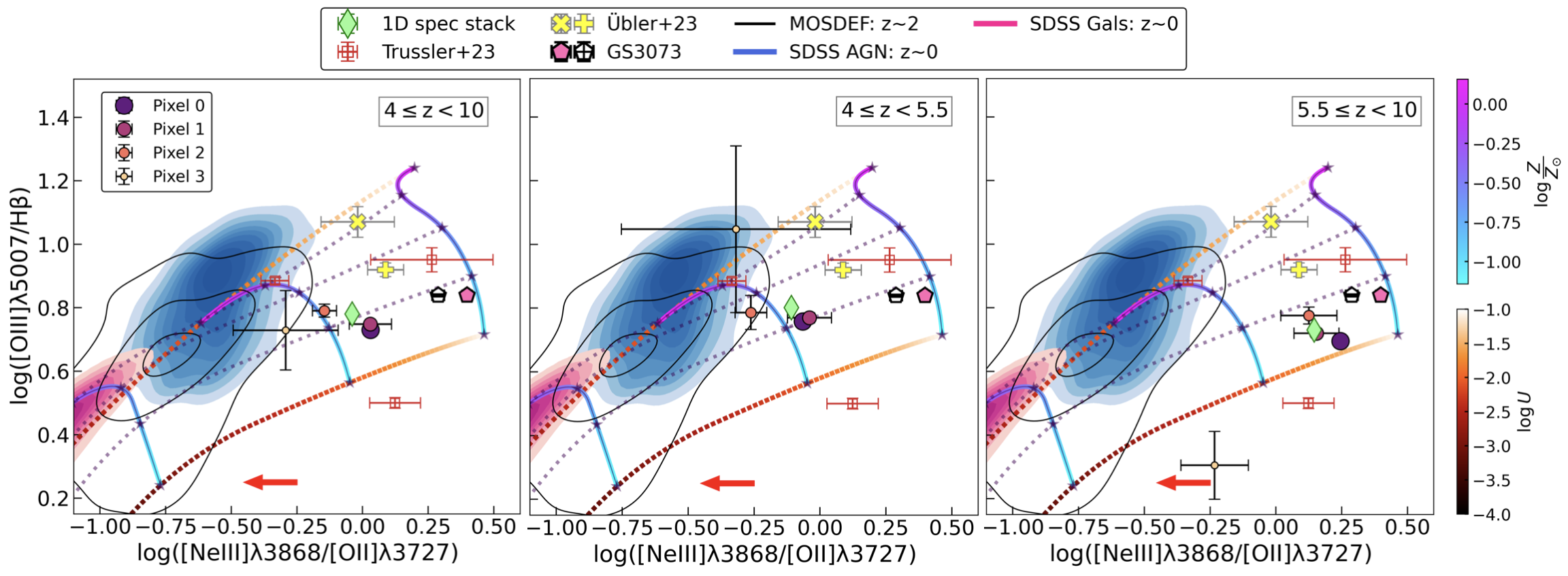}
	\caption{\OIIIL[5007]/\Hbeta--\NeIII/\OIIL[3727] line ratio diagram for the stacked samples. For comparison, we show $z\sim 0$ SDSS AGNs (galaxies) as blue (pink) colormap with contours, $z\sim 2$ MOSDEF galaxies and AGNs (black contours), SMACS~06355, 10612 and 04590 \citetext{red squares; \citealp{trussler2023}; the left-most square of the three is 06355, the type-II AGN identified by \citealp{brinchmann2023}}, the type-I AGN host GS~3073 at $z=5.55$ (filled and hollow pentagon, the latter estimating the flux of \NeIII based on the Case B assumption modulated by the median dust attenuation; Ji~X. in~prep.), and the $z=7.15$ AGN type-I's host galaxy ZS7 \citetext{yellow cross and plus, depending on whether line fluxes are computed from the BLR location or \OIII centroid, respectively; see \citealp{ubler2023} for details}. Spatially resolved ratios for our stacked samples of galaxies at $4 \leq z< 10$, at $4 \leq z< 5.5$, and at $5.5 \leq z< 10$ are plotted as solid circles in all the three panels and are colour and size coded based on the distance (in pixel) from the centre of the galaxy (i.e., pixel 0). Ratios for 1D stacked spectra of our samples are plotted as green rhombs. Overlaid are the star-formation photoionization models of \citet{gutkin2016} 
		at hydrogen densities $\log n{\rm [cm^{-3}]}=2.0$. The dotted and solid coloured lines show the variation of the ionization parameter at fixed metallicity and the variation of metallicity at fixed ionization parameter, respectively (color scales on the right-hand side of the figure). With the spectral resolution of the prism, we measure a blend of \NeIII, \HeIL and \Hzeta; the red horizontal arrow at the bottom of each panel represents the maximum correction for \NeIII/(\NeIII+\HeIL+\Hzeta) of $\approx$0.2 dex (see Sect. \ref{sec:contam}). This correction applies only to our data, and to GS~3073; all other data in the figure have sufficient spectral resolution to deblend \NeIII. }
	\label{fig:models}
\end{figure*}

To better understand the implication of our results, in Fig. \ref{fig:models} we compare them to previous measurements at different redshifts found in the literature, and to photo-ionization models. 
In each of the three panels, the line ratios from the 1D spectrum are intermediate between the spatially resolved line ratios.
Overall, we see radial trends in the three samples, which may be caused by a combination of varying ionization parameter and/or metallicity. In particular, for the $4\leq z < 10$ stack, the radial trend appears to be due mostly to variations in the ionization parameter. However, when we divide the sample at $z=5.5$, we find that the two redshift bins have different average ionization parameter, and the radial variation also shows a component consistent with positive metallicity gradients. This is a direct consequence of the different average value of line ratios seen in Fig. \ref{fig:radial-prof}. This will be extensively discussed in Sect. \ref{sec:disc}.

\begin{table*}[]
	\caption{Radial gradient of $\OIIIL[5007]/\Hbeta$ and $\NeIII/\OIIL[3727]$ ratios, EW$_{\rm \Hbeta}$ and metallicity for the three stacked samples.}
	\centering
	\begin{tabular}{ccc|cccc}
		Sample & Median redshift & \% of galaxies & $\OIIIL[5007]/\Hbeta$ & $\NeIII/\OIIL[3727]$ & EW$_{\rm \Hbeta}$ & $\log(Z/Z_\odot)$\\
		& &    &       [1/pixel]       &       [1/pixel]      & [\AA/pixel]         & [dex/pixel] \\
		\hline
		$4 \leq z<10$ & 5.5 & 63 & 0.024 $\pm$ 0.008 & -0.083 $\pm$ 0.021 & -23 $\pm$ 2 &  0.08 $\pm$ 0.05 \\
		$4 \leq z<5.5$ & 4.7 & 30 & 0.013 $\pm$ 0.014 & -0.085 $\pm$ 0.030 & -20 $\pm$ 2 &  0.07 $\pm$ 0.06 \\
		$5.5 \leq z<10$ & 6.0 & 33 & 0.023 $\pm$ 0.007 & -0.120 $\pm$ 0.033 & -35 $\pm$ 5 &  0.04 $\pm$ 0.05 \\
		\hline
	\end{tabular}
	\label{tab:grad}
	\flushleft 
	\footnotesize {{\bf Notes.} The values reported in this table are the slope of the linear fit of the radial profiles shown in Fig. \ref{fig:radial-prof}, Fig. \ref{fig:EW-stack} and Fig. \ref{fig:radial-prof-metal}. We recall that the pixel size is 0.1", and that the conversion scale (kpc/") used for each sample is 6.113 kpc/", 6.614 kpc/" and 5.832 kpc/", respectively.}
\end{table*}

\subsection{Analysis of \texorpdfstring{\Hbeta}{Hb} equivalent width}
\label{sec:hbeta-EW}

\begin{figure*}
	\centering
	\includegraphics[width=1\linewidth]{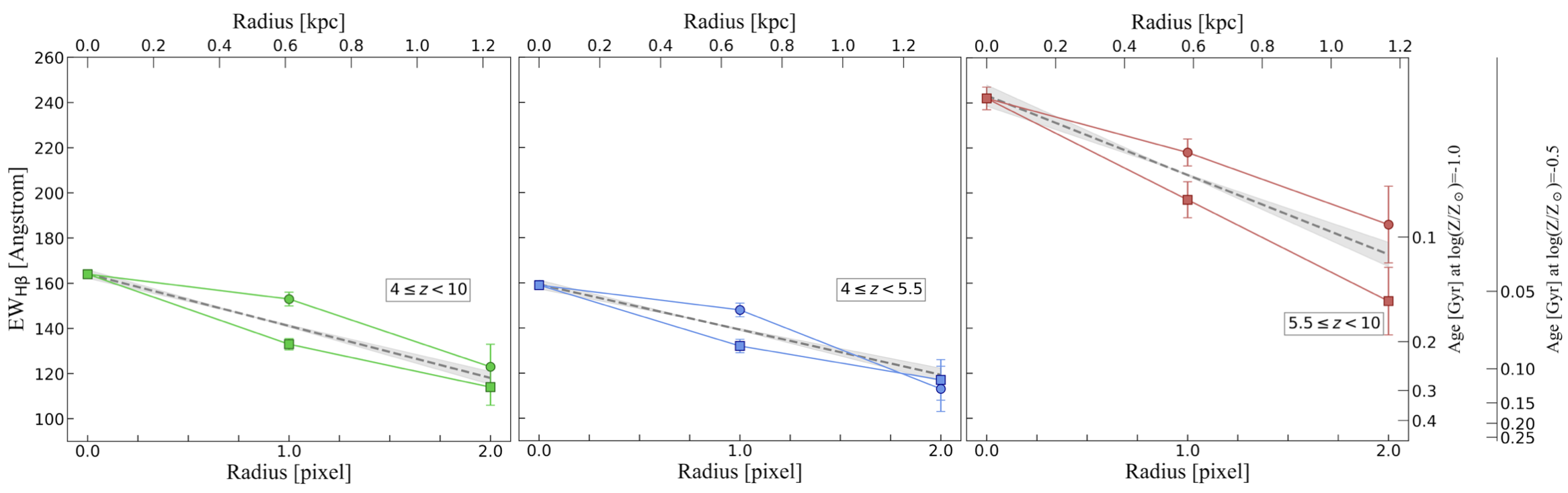}
	\caption{Radial profile of EW$_\Hbeta$ in the stacked samples. From left to right: total stacked sample, stacked sample of $z=4-5.5$ galaxies, stacked sample of $z=5.5-10$ galaxies. The fit for each sample is shown as a dashed line with shadowed region ($1\sigma$ confidence interval). At pixels 1 and 2, different markers are used to discriminate between pixels above (squares) and below (circles) the central pixel 0 in the slit (see also Fig. \ref{fig:2d-spec.b}). Top x-axis as in Fig. \ref{fig:radial-prof}.}
	\label{fig:EW-stack}
\end{figure*}

Fig. \ref{fig:EW-stack} presents the results on the radial profiles of EW$_{\rm \Hbeta}$ for the three samples (full stacked and two z-bins). Overall EW$_{\rm \Hbeta}$ shows a decreasing trend with radius, implying that the emission of \Hbeta is centrally concentrated relative to the continuum, and it is highest in the high-z sample. We use a linear regression to quantify the gradient of EW$_{\rm \Hbeta}$, and the results for the slopes derived for each sample are reported in Tab. \ref{tab:grad}. A similar trend is also found in the radial profile of 4270 at $z=4.023$, shown in Fig. \ref{fig:EW-4270} (image and spectra of 4270 are shown in Fig. \ref{fig:2d-spec}). In this case, we observe a negative radial gradient, steepest in the central regions, followed by a flattening at the outermost radii. 

This rising trend of EW$_{\rm \Hbeta}$ towards the center has some interesting implications on the nature of the innermost regions of these high-z galaxies, depending on the source of ionization. Assuming a star-forming origin, EW$_{\rm \Hbeta}$ is a tracer of the sSFR in SF galaxies, which is directly connected to the age of the stellar population (and indirectly connected to its metallicity, given that a low-metallicity population shows higher EW). The right y-axis of Fig. \ref{fig:EW-stack} shows the scaling between the EW$_{\rm \Hbeta}$ and the age at two fixed different metallicities ($\log(Z/Z_\odot)=-1.0,-0.5$). This scaling has been computed using synthetic spectra calculated with \texttt{Prospector} \citep{johnson2021}, with nebular emission treatments based on \texttt{Cloudy} (see \citealt{byler2017}, for details). We assumed a delayed exponential star-formation history and constant metallicity, and defined the age as the look-back time when the star formation started (we tested different values of the e-folding time $\tau$ between 0.1 and 1~Gyr, but show the results only for a fiducial value of $\tau=0.25$~Gyr). The observed rising trend of EW$_{\rm \Hbeta}$ indicates a negative radial gradient either in metallicity or in age.
In principle, the same trends could be due to central emission boosted by an AGN or by a radial trend in the escape fraction of ionizing photons.
These alternative scenarios will be better discussed in the following section.

\begin{figure*}
	\centering
	\includegraphics[width=1\linewidth]{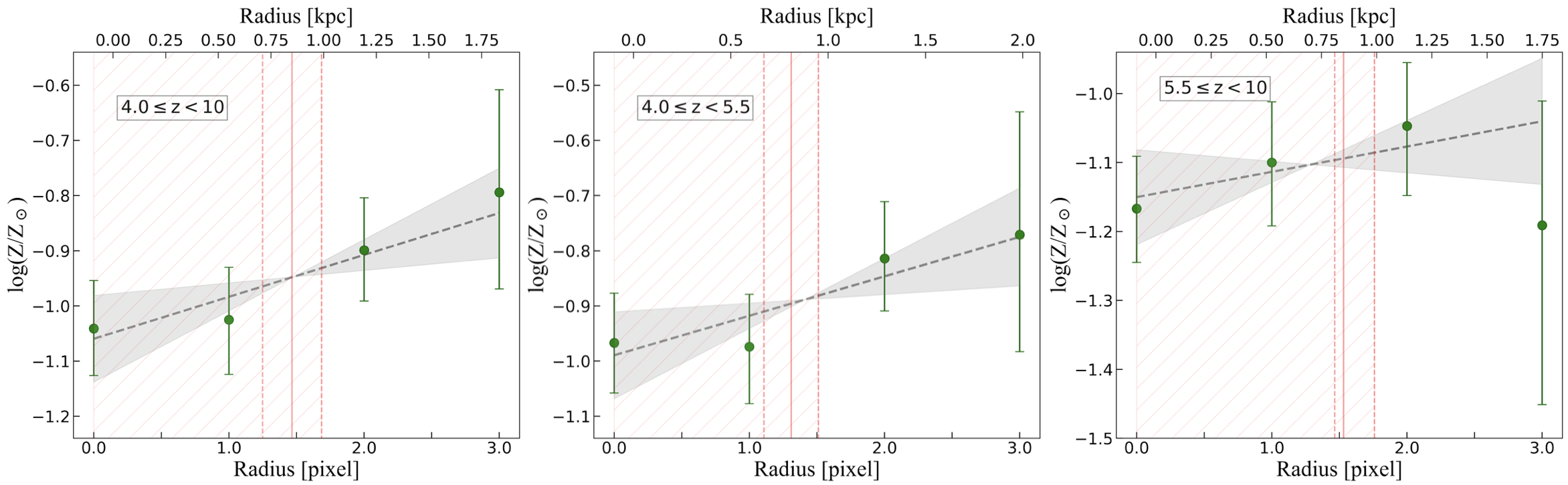}
	\caption{Radial profile of the metallicity in the stacked samples. From left to right: total stacked sample, stacked sample of $z=4-5.5$ galaxies, stacked sample of $z=5.5-10$ galaxies. The fit is shown as a dashed line with shadowed region (1$\sigma$ confidence interval). Orange vertical lines and top x-axis as in Fig. \ref{fig:radial-prof}.}
	\label{fig:radial-prof-metal}
\end{figure*}

\section{Discussion}
\label{sec:disc}


In the former sections, we presented the study of spatially resolved emission line properties in a sample of galaxies at $4\leq z<10$, by stacking the 2D and 1D spectra to get the radial profiles of the line ratios. We also divided this larger sample in two redshift bins ($4\leq z<5.5$ and $5.5\leq z<10$), and we performed the stacking for these two sub-samples. Fig. \ref{fig:radial-prof} and Tab. \ref{tab:grad} show a slightly increasing or flat (for the $4\leq z<5.5$ sub-sample) trend for the $\OIIIL[5007]/\Hbeta$ ratio as a function of radius, while the $\NeIII/\OIIL[3727]$ ratio decreases with radius at more than $3\sigma$ confidence level for both the larger sample and the two sub-samples. 

In Fig. \ref{fig:models}, we compare our results with low-z data in the literature and with star-forming photoionization models. In particular, we show the line ratios measured in the local Universe, both in star-forming galaxies and AGNs \citep[see SDSS DR7 data in pink and blue,][]{kauffmann2003,brinchmann2004,abazajian2004}, in a sample of $\sim 1500$ galaxies and AGNs at $1.37<z<3.8$ \citep[from the MOSDEF survey, black contours][]{kriek2015, reddy2015}, in 3 galaxies at $z>7$ \citep[from the SMACS 0723 \jwst Early Release Observation field, red sqaures,][]{pontoppidan2022,trussler2023}, in the type-I AGN host GS~3073 at $z=5.55$ (filled and hollow pentagon; Ji~X. in~prep.), and in the $z=7.15$ AGN type I's host galaxy ZS7 \citep[yellow cross and plus][]{ubler2023}. For SDSS, we include all the sources at $z>0$ and with S/N$>2$ on the emission lines of interest (\OIIIL[5007], \Hbeta, \OIIL[3727], \NeIII, \Halpha, \NII). It is worth mentioning that SDSS spectra investigate the nuclear region of the galaxies, since they are extracted from the central 3 arcsec of each galaxy. We discriminate between galaxies and AGNs using the BPT diagram \citep{baldwin1981} and considering conservatively as AGNs all the objects lying 0.2~dex above the \citet{Kewley_2001} and \citet{schawinski2007} lines, while as SF galaxies all the objects 0.2~dex below the \citet{kauffmann2003} line. Overplotted are the photoionization models presented in \citet{gutkin2016}, at fixed hydrogen density $\log(n) = 2~\mathrm{cm}^{-3}$. Increasing or decreasing the density by a factor of 10 does not alter the predicted line ratios. The variation of the ionization parameter and the metallicity along the models are highlighted according to the colormaps in the right legend. 

We see two different behaviours for the full stack at $4\leq z < 10$ and for the two redshift bins. For the former, the radial trend appears to be due mostly to variations in ionization parameter. However, when we divide the sample at $z=5.5$, the picture changes.
First, we find that the two redshift bins have different average ionization parameter ($\Delta\sim 0.5$ dex) and metallicity ($\Delta\sim 0.4$ dex), with the highest-redshift bin having the highest average value of $\log\,U$ and the lowest value of $Z$, consistent with other studies \citep{cameron2023,sanders2023,curti2023,nakajima2023}.
Second, the direction of the radial variation of the observed line ratios is intermediate between grid lines of constant $\log\,U$ and constant metallicity.
Note that, in the model grid, we fixed the dust-to-metal mass ratio $\xi_\mathrm{d}=0.3$. Radial variations in $\xi_\mathrm{d}$ would affect both the emission-line ratios we consider, because higher dust depletion would preferentially remove oxygen from the ISM gas (relative to hydrogen and neon).
Specifically, the diagnostic diagram suggests a radial increase in gas metallicity, which could be explained by a radially decreasing $\xi_\mathrm{d}$ at constant metallicity of the ISM (i.e., gas and dust).
However, a decreasing radial gradient of $\xi_\mathrm{d}$ would also increase EW$_{\Hbeta}$, because a lower dust fraction would increase the flux ratio between the UV ionising photons (causing \Hbeta emission) and the rest-optical photons in the continuum near \Hbeta.
Thus, while we cannot rule out a role for radial trends in $\xi_\mathrm{d}$, if such a trend was dominant, it would produce an increasing trend of EW$_{\Hbeta}$ with radius, which is contrary to our observations.

For all the three samples, the radial gradients are in the direction of decreasing ionisation parameter and increasing metallicity, although the large measurement uncertainties prevent us from drawing strong conclusions. For the full stacked sample, the radial trend may be mostly due to the variation of ionization parameter. 
For the two z-bins samples, we find tentative inverted metallicity gradients, and an almost constant ionization parameter as a function of radius, with the only exception being the outermost radius of the higher-z sample for which the ionization parameter drops by $\sim 0.5$ dex. Moreover, the metallicity is on average higher ($\sim 0.4$ below solar) in the $z=4-5.5$ sub-sample than in the higher-z one ($\sim 0.8$ dex below solar), as one would expect for more evolved objects, and is intermediate between the two in the total stacked sample.
We note, however, that this strong metallicity evolution is much larger than what is measured using e.g. strong-line methods \citep[e.g.][]{curti2023}. It is therefore unclear whether the observed increase reflects a physical evolution, or arises from a sample-selection bias. In fact, the highest-z galaxies may have higher average SFR than the targets in the lowest-z bin \citep[see e.g.,][]{scoville2023,tripodi2024,speagle2014,calabro2024,genzel2010}. The fundamental metallicity relation \citep[FMR;][]{mannucci+2010, laralopez2010, cresci2019,baker2023b} predicts that at fixed stellar mass, metallicity decreases with increasing SFR, which could explain the observed trend. The recent observation of deviations from the FMR at $z>6$ \citep{curti2022, curti2023} would only reinforce this trend.

To investigate these inverted metallicity gradients further, we convert both the $\OIIIL[5007]/\Hbeta$ and $\NeIII/\OIIL[3727]$ ratios in O/H, a proxy for gas-phase metallicity, using the calibrations presented in \citet{curti2020}, and updated in \citet{curti2023a,curti2023}. The derived metallicity radial profiles for the three stacked samples are shown in Fig. \ref{fig:radial-prof-metal}. Also in this case, the metallicity in the higher-z sub-sample is -on average- $\sim 0.3$ dex lower than in the $z=4-5.5$ sub-sample. We perform a fit to these profiles using a linear regression, and the results are reported in Tab. \ref{tab:grad}. The increasing trend with radius in all three subsets is suggestive of inverted metallicity gradients, but the large uncertainties on the individual measurements prevent us from drawing definitive conclusions (all three fits are statistically consistent with no gradient).

\begin{figure*}
	\centering
	\includegraphics[width=\textwidth]{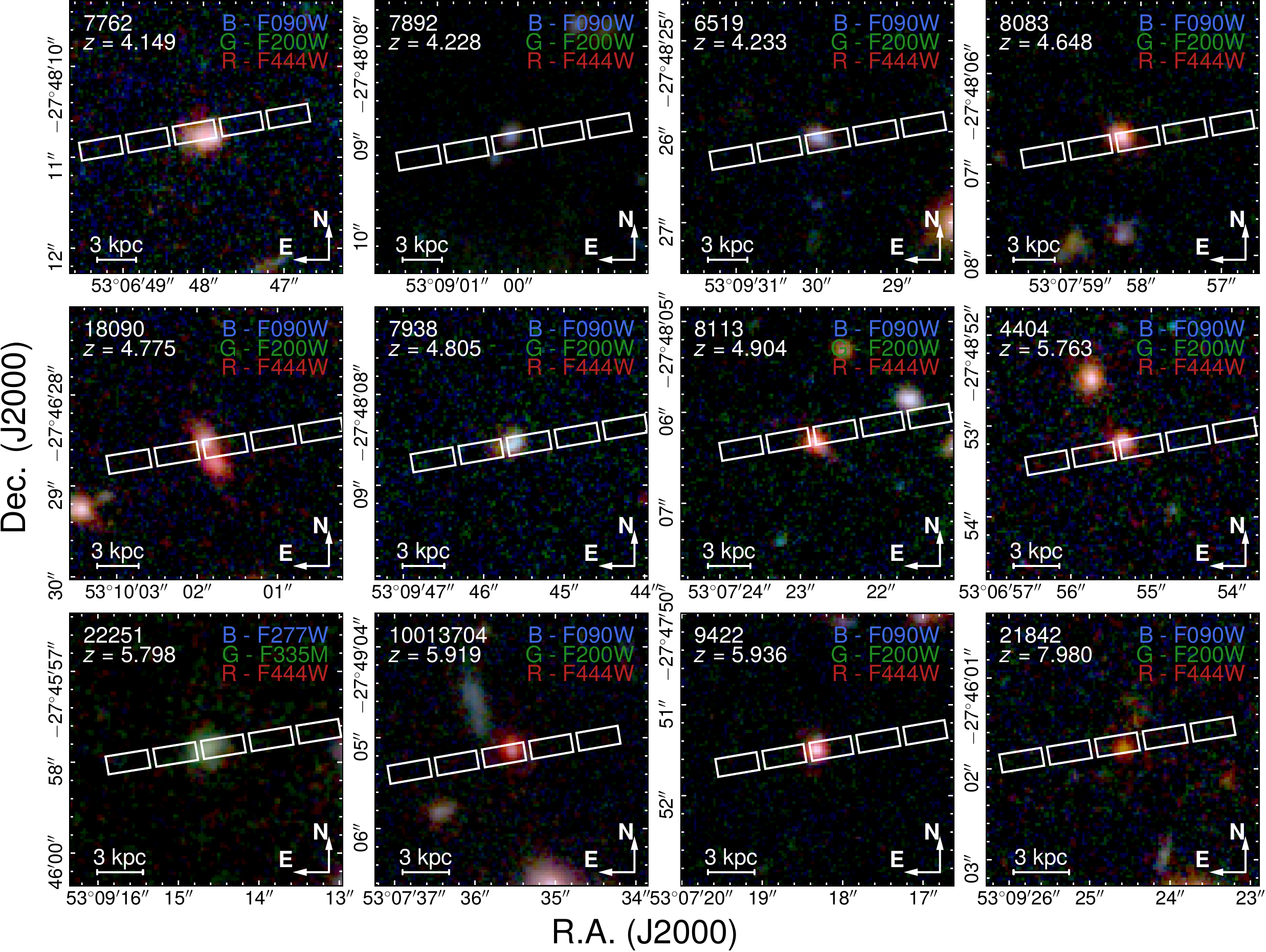}
	\caption{3-arcsec false-colour cutouts of all the galaxies with high S/N and with
		available \jwst/NIRCam photometry, sorted by increasing redshift $z$ (targets 4270 and 16745
		are reported already in Figs.~\ref{fig:2d-spec-alt} and~\ref{fig:2d-spec} and are not repeated
		here).
		The nominal location of the shutters is overlaid; the MSA acquisition accuracy is better than one NIRCam pixel (0.03 arcsec). This figure illustrates the key synergy between NIRCam and NIRSpec/MSA; imaging clarifies that for individual galaxies gradients along the slit may result from complex morphologies. However, because the intra-shutter positions are effectively random between different targets, a stacking analysis over a sufficiently large sample may `marginalise' over peculiar morphologies.
	}\label{fig:postage}
\end{figure*}

In Appendix~\ref{app:indiv}, we consider the radial gradients of five spatially resolved galaxies. However, interpreting the nature of these individual gradients is typically challenging. This complexity arises from the compact nature of galaxies within the considered mass range and redshifts. \citet{degraaff+2023} also found small sizes and complex kinematic features in six $5.5<z<7.4$ galaxies belonging to JADES, three of them showing significant spatial velocity gradient. As a result, individual gradients can be measured only for complex and possibly interacting systems. In this context, the synergy between photometry and spectroscopy is crucial to drawing the correct view about the nature of the gradients (see Fig.~\ref{fig:postage}).
In contrast, when stacking, we `marginalize' over the precision and accuracy of the shutters alignment, making the radial gradients easier to interpret.
Whether a sample of 63 galaxies is sufficient for marginalising over these effects, and enables us to infer the true average gradients of the population, would require simulated observations of realistic galaxies, but this is beyond the scope of the present paper. 

As we have noted, the analysis of the EW of \Hbeta reveals a steep radial profile increasing towards the center in the stacked spectra. Additionally, while analysing the individual galaxy 4270, which shows the clearest gradients among the brightest 5 galaxies in Appendix \ref{app:indiv}, we found a similar trend for EW$_\Hbeta$ vs radius. In SF galaxies, the EW of Balmer lines like \Hbeta and \Halpha tracks the cosmic evolution of the sSFR \citep{fumagalli+2012, sobral+2014}. Indeed, for a fixed age (and, therefore, for a fixed stellar mass-to-light ratio), EW$_\Hbeta$ can be considered as a rough proxy for sSFR. Therefore, the higher EW$_{\Hbeta}$ in the center could be due to our sample being dominated by galaxies undergoing centrally concentrated star formation, leading to the build-up of stellar mass in the core of galaxies (e.g., \citealt{baker2023}; see \citealt{zhang+2012} for a local example in a dwarf galaxy). This phase of `core formation' is indeed expected from both numerical simulations and analytical models \citep{dekel2009, krumholz2018, tacchella2016, tripodi2023, zolotov2015} and observations of high-z galaxies \citep[see e.g., ][]{zhiyuan2023,zhiyuan2023b}, and is fuelled by rates of pristine/low-metallicity gas accretion that are much faster than in the local Universe.
This interpretation would agree with the inverted metallicity gradients, for which the most natural explanation is the continuous dilution of otherwise metal-rich gas in the central regions due to the same gas inflows powering the central star formation.
In principle, the presence of an AGN could also cause a central enhancement of EW$_{\Hbeta}$, given that AGNs tend to have high-EW nebular emission \citep[e.g.,][ for EW$_{\rm \OIII}$]{caccianiga2011}. This may be a plausible scenario for 4270, because the overlap of its line ratios with those of local SDSS AGNs already suggested that 4270 may itself be an AGN host (see Fig. \ref{fig:mod-sing}). 

To ensure that the presence of AGNs did not bias our results on the stacked samples, we excluded from the stacking the 15 AGNs identified in our sample \citep{scholtz2023}, and performed the analysis on the line ratios and EW$_{\Hbeta}$ again. The results are unchanged both for the total and the two z-bin stacked samples. That is, even excluding the AGNs, we find flat radial profiles for \OIIIL[5007]/\Hbeta, and declining for \NeIII/\OIIL[3727], and a steep increase of EW$_{\Hbeta}$ towards the center. 
This suggests that AGNs do not explain the observed trends in EW$_{\Hbeta}$, which, therefore, must be due to stellar-population trends.

It should also be noted that the redshift range we are probing is relatively uncharted for this topic, therefore our interpretation is uncertain. In particular, star-forming galaxies in burst phases are much more common in the young Universe \citep{ciesla2023,dressler2023, looser2023, endsley2023}, as also expected from models \citep[see e.g., ][]{faucher2018,tacchella2020}. These objects have much higher EW$_\Hbeta$ and EW$_{\rm \OIII}$ than star-forming galaxies in the local Universe \citep[e.g.,][]{smit+2014,vanderwel2011,maseda2014,boyett2024}, therefore they may equal or even exceed the EWs of AGNs \citep[for type-2 AGN, EW$_{\OIII}<500$~\AA; e.g.,][]{caccianiga2011}. In principle, the observed trends could be caused by the fact that, on average, EW$_\Hbeta$ increases with z \citep[e.g.,][]{smit+2014}, whereas the typical size of galaxies decreases \citep{vanderwel2014,ji2024}. However, we still observe the same trend between EW$_\Hbeta$ and radius even after splitting our sample in two redshift bins, which suggests that the trend is real and not due to the observational conspiracy between EW$_\Hbeta$, size and redshift.
Alternatively, if starburst galaxies had both higher EW$_\Hbeta$ and smaller size than non-starburst galaxies, this could also produce the decreasing radial gradients of EW$_\Hbeta$ in the stack.
This is because, under this hypothesis, starburst galaxies (with their high EW$_\Hbeta$) would dominate the central pixels of the stack, while more extended, less bursty galaxies (with lower EW$_\Hbeta$) would dominate the outer pixels of the stack, thus creating a spurious gradient.
However, distinguishing between inverted gradients and correlations between galaxy structure and emission-line ratios (and EW$_\Hbeta$) requires further dividing the sample; for this technique to be effective, we need a larger sample than the currently available one.
NIRCam could help evaluate this possibility, combining SED fitting (to identify starbursts) and NIRCam medium- vs wide-band imaging (to measure EW gradients).

Alternatively, IFU observations with NIRSpec can directly measure spectral gradients in galaxies \citep{rodriguez-delpino+2023,arribas+2023,venturi2024}. However, NIRSpec/integral field spectroscopy (IFS) is best suited for large, bright galaxies, primarily due to its low multiplexity. Using this instrument configuration for targeting a large sample of low-mass, high-z galaxies with a depth comparable to JADES, would require an inordinate amount of time.
This is compounded by the lower instrument sensitivity and, possibly, by a larger PSF at 1--3~\mum \citep[at least in the current cube reconstruction; e.g.,][]{deugenio2023}.
For all these reasons, we deem it unlikely that NIRSpec/IFS can be competitive for building large samples -- at least in the redshift and mass range explored in this article.

\section{Summary and Conclusions}

We study the radial profiles of the $\OIIIL[5007]/\Hbeta$ and $\NeIII/\OIIall$ line ratios, and of the EW$_{\rm \Hbeta}$ in a sample of 63 galaxies at $4\leq z<10$, using publicly available \jwst/NIRSpec data from JADES. We focus on deep, spatially resolved spectroscopy in the GOODS-S extragalactic field, since we want to study spatially resolved emission line gradients. Our findings are summarized as follows.
\begin{itemize}
	\item By performing an initial selection of 19 galaxies with S/N$>5$ on the lines of interest, we find that 5 of them show flat radial trends of the line ratios. Specifically, for the two galaxies 4270 at $z=4.023$ and 16745 at $z=5.5666$, these trends are detected or marginally detected up to $\sim 2$ kpc from the center of the galaxy. These can be caused by merging events or re-accretion of metal enriched gas. However, the complex morphology shown in the NIRCam images makes the interpretation of gradients arduous. This highlights the importance of the synergy between imaging and spectroscopy for capturing the physical properties of these objects.  
	\item  To increase the S/N ratio on the lines of interest and to marginalize over the complex galaxy morphologies, we stack the whole sample of 63 galaxies and two sub-samples, created by dividing the larger sample in two redshift bins ($4\leq z<5.5$, and $5.5\leq z<10$). We found a slightly increasing or flat radial trend for $\OIIIL[5007]/\Hbeta$, while a decreasing radial trend for $\NeIII/\OIIall$ in all the three stacked samples.
	\item By comparing with photoionization models, we find that the two redshift bins have different average ionization parameters ($\Delta \sim 0.5$ dex, increasing with redshift), with tentative evidence of an inverted metallicity gradient. The metallicity is on average $\sim 0.3$ dex higher in the $z=4-5.5$ sub-sample, implying an evolution stronger than measured using e.g., strong-line methods. This is still valid if adopting the calibrations presented in \citet{curti2020} to convert line ratios to metallicity. It remains difficult to discriminate whether this reflects a physical evolution or a selection bias, since galaxies with higher SFR at high-z are expected to have lower metallicity at fixed stellar mass. 
	\item We also find negative radial gradients for EW$_\Hbeta$, even when removing known AGNs from the sample. This may indicate that -- on average --
	our sample of high-z galaxies is dominated by central starbursts, fuelled by high mass accretion rates. The latter could also explain the inverted metallicity gradients. 
\end{itemize}

This work led to the first tentative detection of metallicity gradients and of negative gradient of EW$_{\Hbeta}$ in a stacked sample of $4\leq z<10$ galaxies. There are several possible avenues to confirm and refine this study. For galaxies with high-EW emission lines, we could exploit the spatial resolution of NIRCam and the excellent wide- and medium-filter coverage of the JADES Origins Field \citep{eisenstein+2023b}.
NIRCam slitless spectroscopy also offers promising possibilities \citetext{e.g. FRESCO, \citealp{oesch2023}; CONGRESS, (GO-3577), Sun et~al. in~prep}.
For fainter and/or low-EW targets, and especially regarding the study of metallicity gradients, we need larger samples of comparable depth.
Finally, for the highest-mass galaxies, which are typically the most extended, we would require a different approach to the background subtraction (to avoid self subtraction) or to the observations \citetext{e.g., the NIRSpec/IFS, \citealp{boker+2022,rodriguez-delpino+2023,arribas+2023}}.

\vspace{1cm}
\noindent \textit{Acknowledgment.} RT and MB acknowledge support from the ERC Grant FIRSTLIGHT and Slovenian national research agency ARIS through grants N1-0238 and P1-0188. RT acknowledges financial support from the University of Trieste. RT acknowledges support from PRIN MIUR project “Black Hole winds and the Baryon Life Cycle of Galaxies: the stone-guest at the galaxy evolution supper”, contract \#2017PH3WAT.
FDE, RM, JS, WB, XJ and JW acknowledge support by the Science and Technology Facilities Council (STFC), by the ERC through Advanced Grant 695671 ``QUENCH'', and by the
UKRI Frontier Research grant RISEandFALL. RM also acknowledges funding from a research professorship from the Royal Society.
MC acknowledges the support from the ESO Fellowship Programme. 
AJB and AJC acknowledge funding from the ``First Galaxies'' Advanced Grant from the European Research Council (ERC) under the European Union’s Horizon 2020 research and innovation program (Grant agreement No. 789056). 
JT acknowledges support by the ERC Advanced Investigator Grant EPOCHS (788113) from the European Research Council (ERC) (PI Conselice).
SA acknowledges support from Grant PID2021-127718NB-I00 funded by the Spanish Ministry of Science and Innovation/State Agency of Research (MICIN/AEI/ 10.13039/501100011033).
SC and GV acknowledge support by European Union's HE ERC Starting Grant No. 101040227 - WINGS.
ZJ and CNAW acknowledge funding from JWST/NIRCam contract to the University of Arizona NAS5-02015.
BER acknowledges support from the NIRCam Science Team contract to the University of Arizona, NAS5-02015, and JWST Program 3215.
H{\"U} gratefully acknowledges support by the Isaac Newton Trust and by the Kavli Foundation through a Newton-Kavli Junior Fellowship.
The authors acknowledge use of the lux supercomputer at UC Santa Cruz, funded by NSF MRI grant AST 1828315.
This work is based on observations made with the NASA/ESA/CSA James Webb Space Telescope. The data were obtained from the \href{https://mast.stsci.edu/portal/Mashup/Clients/Mast/Portal.html}{Mikulski Archive for Space Telescopes (MAST)} at the Space Telescope Science Institute, which is operated by the Association of Universities for Research in Astronomy, Inc., under NASA contract NAS 5-03127 for JWST. These observations are associated with programme PID~1210 (NIRCam-NIRSpec galaxy assembly survey - GOODS-S - part \#1b; PI~N.~L\"utzgendorf).
Reduced and calibrated images and spectra were obtained from the \href{https://jades-survey.github.io/}{JADES Collaboration}, via the high-level science products page on \href{https://archive.stsci.edu/hlsp/jades}{MAST.}

\textit{Facilities:} \jwst. \textit{Software:} astropy \citep{astropy2022}, Matplotlib \citep{matplotlib2007}, SciPy \citep{scipy2023}, emcee \citep{foreman2013}.

\bibliography{biblio}{}
\bibliographystyle{aa}

\appendix
\section{Assessing the spatial resolution of the data}
\label{app:PSF}

A key requirement for making meaningful gradient measurements is that the data do contain spatial information. In this section, we show that the data do contain marginally resolved spatial information, by comparing the spatial profile of the \OIIIL[5007] nebular emission with the model point-spread function (PSF).

\begin{figure}
    \centering
    \includegraphics[width=0.95\linewidth]{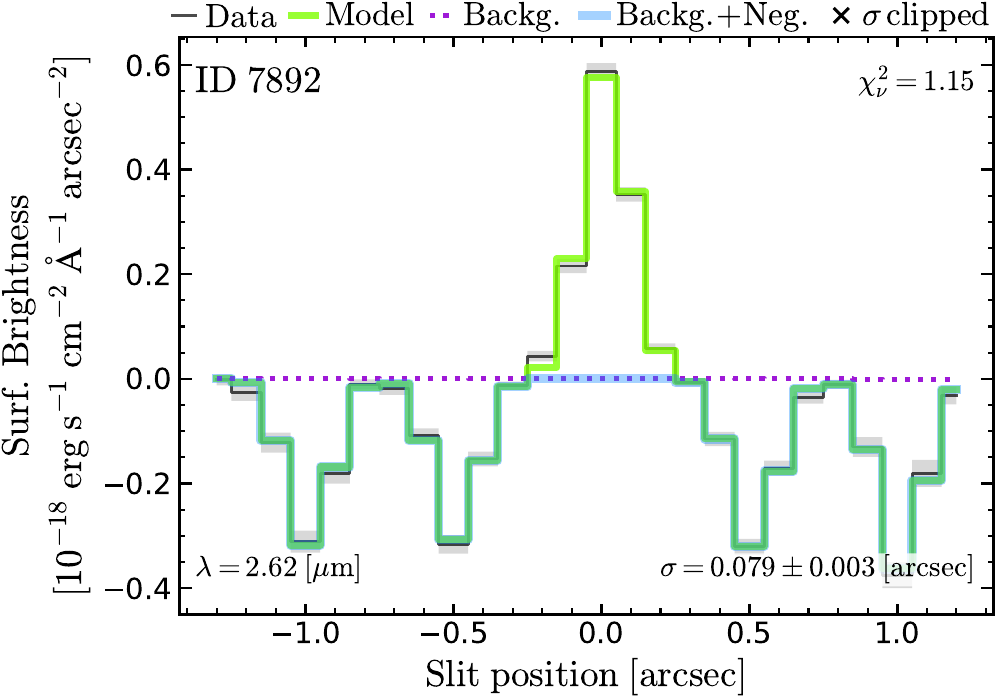}
    \caption{Example of model fit to the galaxy spatial profile, for galaxy ID 7892. We show the data (black),
    the model background (dotted red), the background plus the negative traces (blue), and the best-fit model (green). The dark green colour is due to overlapping blue and green. We report the observed wavelength of \OIIIL[5007] (bottom left) and the best-fit $\sigma$ (bottom right).} 
    \label{fig:profile_fit}
\end{figure}
\begin{figure}
    \centering
    \includegraphics[width=0.95\linewidth]{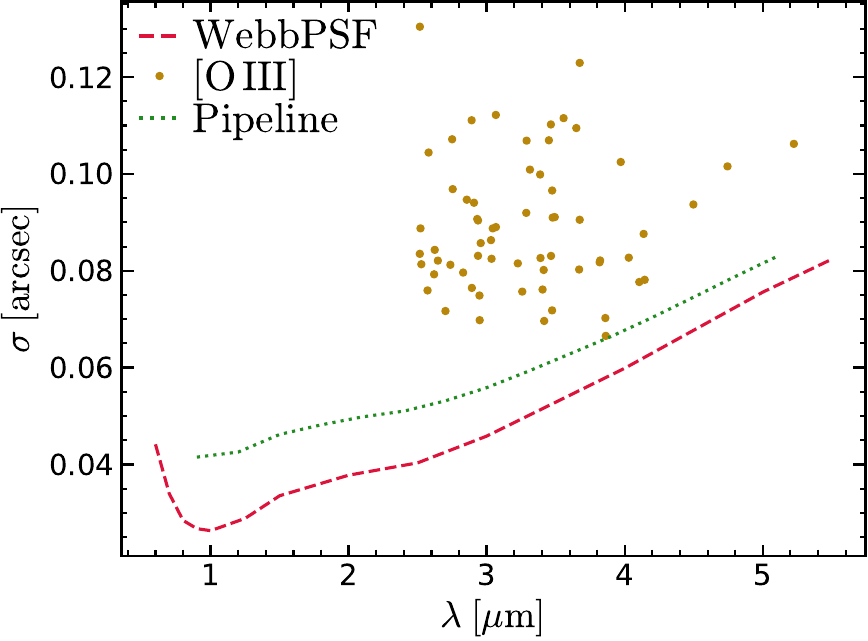}
    \caption{Comparison of the radial extent of our data (using \OIIIL[5007], circles) to the PSF model frow \texttt{webbpsf} (dashed line) and to the model from the data reduction pipeline (dotted line). The galaxies in our sample extend beyond the PSF $\sigma$.}
    \label{fig:psf}
\end{figure}

For the galaxies, we measure the light profile along the slit by summing the 2D spectrum in a window of three wavelength pixels, centred on the \OIIIL[5007] wavelength. As a model, we used a linear background (two free parameters), a Gaussian representing the source profile (three free parameters),
and four more Gaussians representing the negative source traces, which arise from the nod sky subtraction \citetext{Sect.~\ref{sec:obs} and \citealp{bunker+2023b}}.
These four negative Gaussians add nine free parameters in total, as we discuss below. The centroid of each of the four negative Gaussians is free (four out of nine parameters), but is subject to a Gaussian probability prior placing it one or two shutters away from the source (prior dispersion 0.2~arcsec). The flux of each of the four negative Gaussians is also free (four out of nine parameters), but again is subject to being smaller (in absolute value) than the flux of the source; in addition, we place a Gaussian probability prior such that the negative traces have flux which is within a fraction of the source flux (with prior dispersion 0.5). The final free parameter for the
negative traces is their common dispersion, which is constrained to be identical to that of the source within a Gaussian probability prior with fractional dispersion 0.5. The simultaneous fitting of the negative traces is crucial for a reliable estimate of
the galaxy's profile in the regime where the PSF is mildly under-sampled. This model is optimised using again a MCMC integrator, and our fiducial value of the source extent is the $\sigma$ of the source Gaussian, for which we quote the 50\textsuperscript{th} percentile as fiducial value, and half the
16\textsuperscript{th}--84\textsuperscript{th} inter-percentile range as uncertainty.
An example galaxy profile is reported in Fig.~\ref{fig:profile_fit}.
The $\sigma$ of each galaxy in our sample is shown as a dot in Fig~\ref{fig:psf}.

For the MSA PSF, we use the model from \href{https://github.com/spacetelescope/webbpsf}{\texttt{webbpsf}}, for which we report the Gaussian-equivalent $\sigma$, defined by
$\sigma \equiv \mathrm{FWHM}/\sqrt{\ln{256}}$ (dashed line in Fig.~\ref{fig:psf}).
We tested that following the approach
of \citet{degraaff+2023} gives consistent results.

We note that the process of co-adding the 2D spectra in the data reduction pipeline does not align the traces to the sub-pixel level. This is done to avoid resampling the data, which increases the correlation between the pixels. The pipeline currently aligns the 2D spectra from each visit by finding their peak. To model the effect of this procedure, we create mock 2D spectra with the wavelength-dependent $\sigma$ of \texttt{webbpsf}, but for each realisation, we sum 20 exposures with an added random offset of $\pm0.5$~pixel, or 0.05~arcsec.
We then fit the spatial profile of this distribution, and find that indeed its profile is broader than the model PSF (dotted line in Fig.~\ref{fig:psf}), but still below the values measured for our data.  By comparing the dotted line with the dashed line representing \texttt{webbpsf}, we estimate the loss in resolution from the alignment procedure to be equivalent to convolving the native PSF with a Gaussian PSF having $\sigma=0.03$~arcsec.
\section{Analysis of individual sources}
\label{app:indiv}

\begin{figure*}
    \centering
    \includegraphics[width=0.95\linewidth]{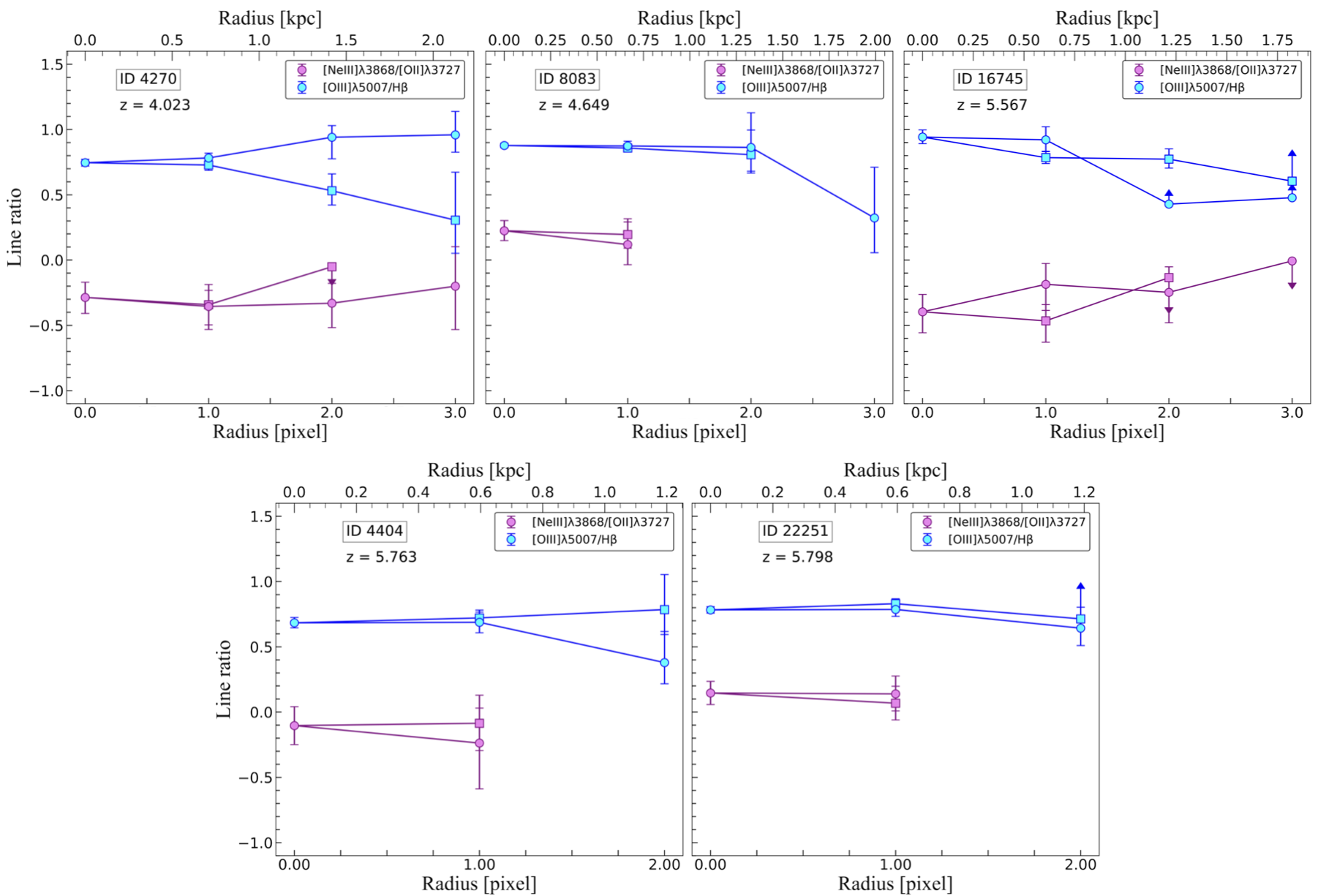}
    \caption{Five best cases of radial profiles of line ratios. Cyan (violet) dots mark the logarithm of $\OIIIL[5007]$/\Hbeta ($\NeIII/\OIIL[3727]$) ratio as a function of the radius expressed in pixels (pixel 0 is the centre of the galaxy). Different markers are used to discriminate between pixels above (squares) and below (circles) the central pixel 0 in the slit (see also Fig. \ref{fig:2d-spec.b}). The ID of each object and the redshift are specified in the top left corner of each plot.}
    \label{fig:rad-prof-best5}
\end{figure*}


In order to perform a spatially resolved study of the line ratios \OIIIL[5007]/\Hbeta and \NeIII/\OIIL[3727] in a sample of galaxies at $4\leq z<10$, we initially select all the publicly available galaxies belonging to the JADES survey with S/N$>=5$ on the flux of each spectral line\footnote{These fluxes are also publicly available, and are computed from the line fitting of the 1D spectrum of each source \citep{bunker+2023b}.} at $4\leq z<10$: we find that 19 objects satisfy these criteria. 

Fig. \ref{fig:2d-spec-alt.b} shows the 2D spectrum of one of those objects (ID 4270 at $z=4.023$), where the y-axis has been shifted with respect to the brightest pixel along the slit, which represents the centre of the galaxy. We shift all the 2D spectra according to the centre of each galaxy. The \OIIIL[5007] emission of 4270 is very bright and clearly extended over 7 pixels, that is $\sim 5$ kpc at the galaxy redshift. We extract the spectra from these 7 pixels (from pixel -3 to +3). For this object, $\NeIII$ is blended with $\NeIIIb$, and we take this into account when fitting the spectrum (see Sect. \ref{sec:fitting})

Another interesting object is 16745, which also has a very bright \OIIIL[5007] emission, extended over 7 pixels that corresponds to $\sim 4$ kpc at the galaxy redshift (see Figs. \ref{fig:2d-spec.b} and \ref{fig:2d-spec.c} for 2D and 1D spectra). \citet{degraaff+2023} found disc-like kinematics in 16745, with $v/\sigma\sim 2$. We fitted both pairs of emission line as above, considering that in this case there is no strong blending between the $\NeIIIall$ doublet.

Top left and top right panels of Fig. \ref{fig:rad-prof-best5} present the radial profile of the ratios between the integrated fluxes of \OIIIL[5007] and \Hbeta (cyan dots), and between \NeIII and \OIIL[3727] (violet dots) for 4270 and 16745 along with other three bright sources. Since the choice of the pixel sign is arbitrary, we show the fluxes as a function of the absolute value of the radius. The interpretation of these profiles is challenging, since the morphology of these sources is complex (see Figs. \ref{fig:2d-spec-alt.a} and \ref{fig:2d-spec.a}), probably caused by the presence of a companion or satellite galaxy.

We apply the same method for each object in this sample, always taking into account the blending of the \OIIIall doublet, and the blending of the \NeIIIall doublet.  The results for the other 4 best cases are shown in the other panels of Fig. \ref{fig:rad-prof-best5}, where the object ID and redshift are specified in the top left corner of each plot. We select the cases in which the \OIIIL[5007]/\Hbeta ratio is detected up to pixel 2 at least, and \NeIII/\OIIL[3727] up to pixel 1 at least. The other sources present upper/lower-limits or non-detections in the off-center pixels, especially for the $\NeIII/\OIIL[3727]$ ratio.

\begin{figure}
    \centering
    \includegraphics[width=0.95\linewidth]{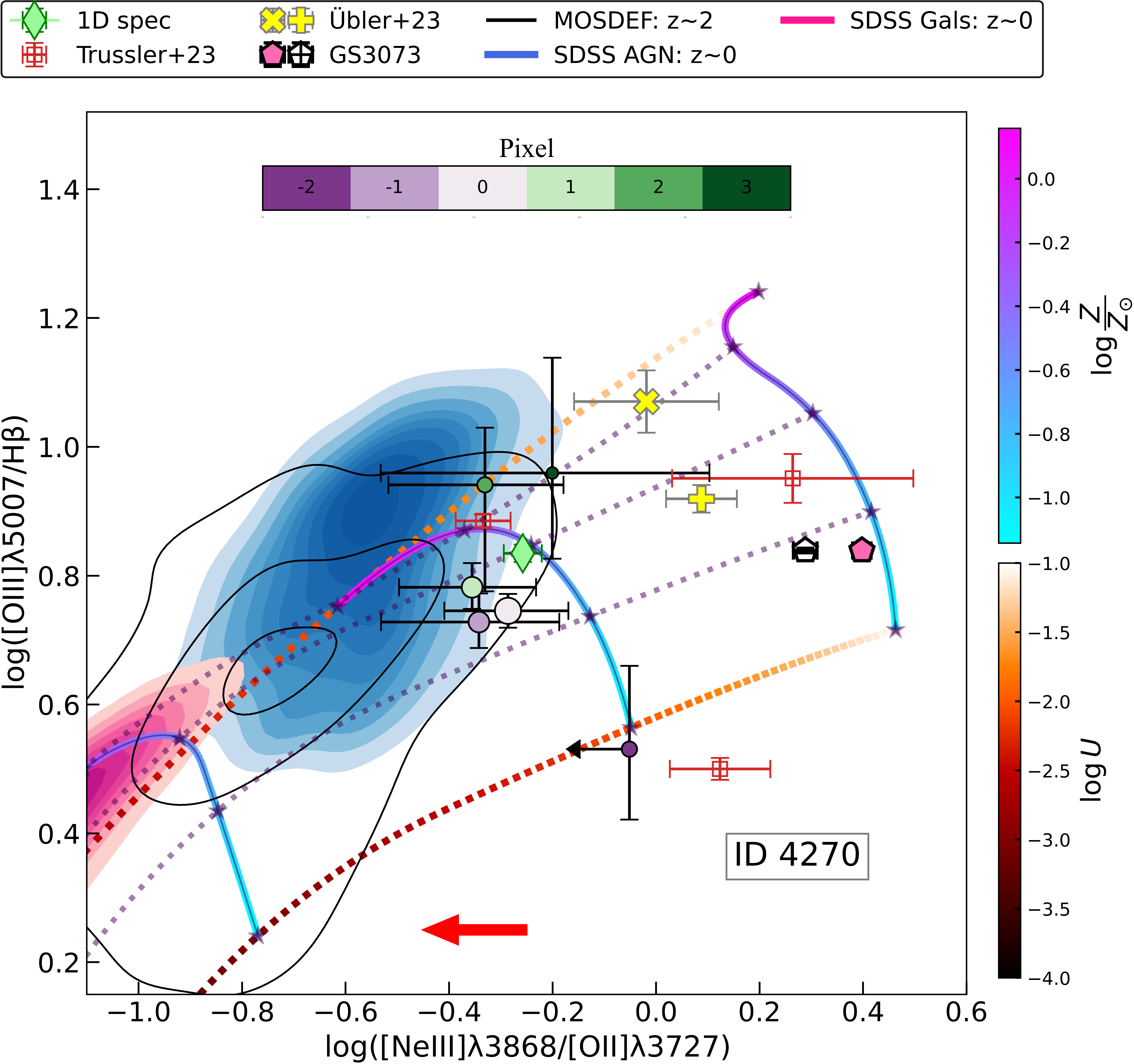}
    \caption{\OIIIL[5007]/\Hbeta--\NeIII/\OIIL[3727] line ratio diagram for ID~4270. Ratios for 4270 at $z=4.023$ are plotted as solid circles, that are colour and size coded based on the distance from the centre of the galaxy (i.e., pixel 0 is the lightest and biggest). For comparison, we show $z\sim 0$ SDSS AGNs (galaxies) as blue (pink) colormap with contours, $z\sim 2$ MOSDEF galaxies and AGNs (black contours), SMACS~06355, 10612 and 04590 \citetext{red squares; \citealp{trussler2023}; the left-most square of the three is 06355, the type-II AGN identified by \citealp{brinchmann2023}}, the type-I AGN host GS~3073 at $z=5.55$ (filled and hollow pentagon; Ji~X. in~prep.), and the $z=7.15$ AGN type-I's host galaxy ZS7 \citetext{yellow cross and plus, depending on whether line fluxes are computed from the BLR location or \OIII centroid, respectively; see \citealp{ubler2023} for details}. Spatially resolved ratios for our stacked samples of galaxies at $4 \leq z< 10$, at $4 \leq z< 5.5$, and at $5.5 \leq z< 10$ are plotted as solid circles in all the three panels and are colour and size coded based on the distance (in pixel) from the centre of the galaxy (i.e., pixel 0). Ratios for 1D stacked spectra of our samples are plotted as green rhombs. Overlaid are the star-formation photoionization models of \citet{gutkin2016} 
    at hydrogen densities $\log n{\rm [cm^{-3}]}=2.0$. The dotted and solid coloured lines show the variation of the ionization parameter at fixed metallicity and the variation of metallicity at fixed ionization parameter, respectively (color scales on the right-hand side of the figure). With the spectral resolution of the prism, we measure a blend of \NeIII, \HeIL and \Hzeta; the red horizontal arrow at the bottom of each panel represents the maximum correction for \NeIII/(\NeIII+\HeIL+\Hzeta) of $\approx$0.2 dex (see Sect. \ref{sec:contam}). This correction applies only to our data, and to GS~3073; all other data in the figure have sufficient spectral resolution to deblend \NeIII.}
    \label{fig:mod-sing}
\end{figure}

Overall, the most extended target is found to be 4270, which shows emission of \NeIII and \OIIL[3727] up to pixel 3 (i.e., $\sim 2$ kpc from the galaxy centre). Fig. \ref{fig:mod-sing} compares our emission-line ratios for the 2D and 1D spectra to the literature. When comparing with low-z data in the literature and with the star-forming photoionization models of \citet{gutkin2016}, we see that 4270 is consistent with the redshift evolution of the \OIIIL[5007]/\Hbeta and \NeIII/\OIIL[3727] ratios, which both increase from $z=0$ to $z\sim2$ and $z\sim7$. In particular, the measurements for
this target overlap with the high-\NeIII/\OIIL[3727] envelope of the SDSS AGNs. The spatially resolved measurements have large uncertainties, yet they show some tentative dependencies on metallicity and ionization parameter. We can see that the ratios of the 1D spectrum are averages between the resolved measurements, which is a reassuring consistency check, implying that the 1D spectrum captures the average characteristics of the galaxy.

\begin{figure}
    \centering
    \includegraphics[width=0.95\linewidth]{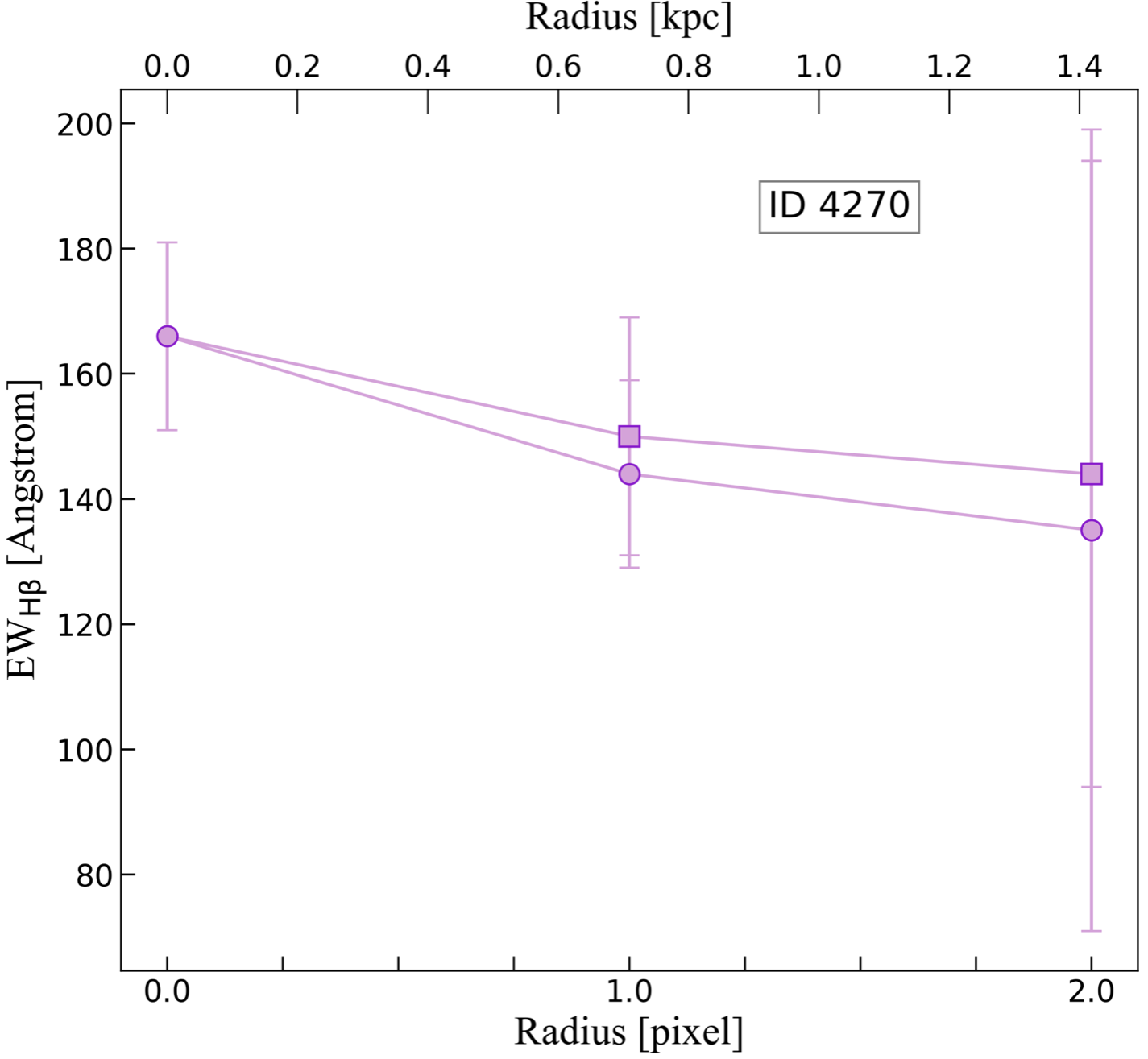}
    \caption{Radial profile of \Hbeta EW in 4270 at $z=4.023$. At pixels 1 and 2, different markers are used to discriminate between pixels above (squares) and below (circles) the central pixel 0 in the slit (see also Fig. \ref{fig:2d-spec.b}).}
    \label{fig:EW-4270}
\end{figure}

The flat shape of these radial profiles can be caused by merging events or re-accretion of metal-enriched gas \citep{oppenheimer2008,ubler2014}. Specifically for the cases of 4270 at $z=4.023$ and 16745 at $z=5.566$, both these scenarios may be likely. This is because both sources have a complex morphology (Figs.~\ref{fig:2d-spec-alt.a} and \ref{fig:2d-spec.a}) and it is unclear whether we are observing satellite galaxies or simply a clumpy
morphology. The position of the shutters is such that we may trace some part of the companion emission, and the central pixel (pixel 0) is actually tracing part of the central emission, more than the actual centre of the source. Moreover, note that the position of the shutters is known with finite precision. As a word of caution, complex morphology and slit position make the interpretation of gradients in individual sources really challenging, especially when the position of the shutters is not aligned with the source.
As we have seen, the synergy between slit spectroscopy and imaging is of critical importance to grasp the complex morphology and physical properties of high-redshift galaxies; 2D maps of the stellar mass and/or gas kinematics are critical to understanding the structure of individual sources, and to correctly interpreting the observed spatial variations.
In Fig. \ref{fig:mod-sing}, we compare the results for 4270 with other galaxies and AGNs at different redshifts from the literature. Interestingly, the line ratios of 4270 overlap with the SDSS AGNs, suggesting that this object may host an AGN. Applying the \Hzeta+\HeIL contamination correction enhances
the overlap with SDSS AGN, and would not change our conclusions. 
When comparing with photoionization models, at fixed hydrogen density ($\log n{\rm [cm^{-3}]}=2.0$), the metallicity is almost solar at the galaxy centre and increases up to 0.3 dex with radius. This inverted gradient can be explained by recent episodes of pristine gas accretion to the galaxy centre or strong radial flows, and it has been previously found in galaxies at lower redshift, $1.2<z<2.5$ \citep{curti2020}. There also seems to be a trend with ionization parameter, with higher ionization towards larger radii, although
the measurement uncertainties are large. Indeed, since the major
photoionization sources (e.g., stars, AGN) are typically concentrated in the centre, this scenario of higher ionization in the outskirts appears unlikely, even if considering the presence of a companion. Therefore, we conservatively conclude that the ionization parameter radial profile is consistent with flat, given the large errorbars along the x-axis. For this target, we also find that EW$_\Hbeta$ decreases with radius (Fig.~\ref{fig:EW-4270}), which would be in agreement with both the AGN scenario
as well as a central starburst, fuelled by inflow of relatively metal-poor gas.

\end{document}